%% file: Main.tex
\documentclass[acmsmall]{acmart}

\usepackage{amsmath,amsfonts}
 
\usepackage{amssymb}

\usepackage{wrapfig} % 浮动表格
\usepackage{bbding} %对勾叉号

\usepackage{lipsum}
\usepackage{tikz}
\usetikzlibrary{arrows.meta}
\usetikzlibrary{automata,positioning}

\usepackage{threeparttable}

\definecolor{tabred}{RGB}{230,36,0}%
\definecolor{tabgreen}{RGB}{0,116,21}%
\definecolor{taborange}{RGB}{250,124,30}%
\definecolor{tabbrown}{RGB}{171,70,0}%
\definecolor{tabyellow}{RGB}{251,253,169}%
\newcommand*{\vcorr}{%
  \vadjust{\vspace{-\dp\csname @arstrutbox\endcsname}}%
  \global\let\vcorr\relax
}% 

\usepackage{mathtools}

\usepackage{lscape}
\usepackage[figuresright]{rotating}
\usepackage[graphicx]{realboxes}
\usepackage{adjustbox}
\usepackage{caption}

\usepackage{subfigure}
\usepackage{caption}

\usepackage{colortbl} 
\usepackage{xfrac}

\usepackage{appendix}
\usepackage{float}

\usepackage{fbox} %use box
\usepackage{fancybox}%use other types of boxes
\usepackage{framed}

\usepackage{multirow}

\def\BibTeX{{\rm B\kern-.05em{\sc i\kern-.025em b}\kern-.08em
    T\kern-.1667em\lower.7ex\hbox{E}\kern-.125emX}}
    
\usepackage{CJKutf8}  

\usepackage{hyperref}

\usepackage{tabularx,makecell, caption}
\usepackage{enumitem} % to control Itemization spacing

\newcolumntype{L}{>{\arraybackslash}X}

\usepackage{multicol}
\usepackage{listings}
\usepackage{blindtext}
\usepackage{etoolbox,xstring,mfirstuc,textcase}
\usepackage{listings}

\usepackage{subfiles}
\usepackage[most]{tcolorbox}

\usepackage{xcolor}
\theoremstyle{plain}                
\theoremstyle{definition}       

\usepackage{siunitx}
\usepackage{comment}
\usepackage{indentfirst} 
\usepackage{framed} 

\usepackage[font=small,labelfont=bf,tableposition=top]{caption}
\usepackage{booktabs}
\usepackage{dashbox}

\usepackage{listings}
\usepackage{url}

\lstdefinelanguage[ECMAScript2015]{JavaScript}[]{JavaScript}{
basicstyle=\tiny,
  morekeywords=[1]{await, async, case, catch, class, const, default, do,
    enum, export, extends, finally, from, implements, import, instanceof,
    let, static, super, switch, throw, try},
  morestring=[b]` % Interpolation strings.
}

\lstdefinelanguage{JavaScript}{
basicstyle=\tiny,
  morekeywords=[1]{break, continue, delete, else, for, function, if, in,
    new, return, this, typeof, var, void, while, with},
  morekeywords=[2]{false, null, true, boolean, number, undefined,
    Array, Boolean, Date, Math, Number, String, Object},
  morekeywords=[3]{eval, parseInt, parseFloat, escape, unescape},
  sensitive,
  morecomment=[s]{/*}{*/},
  morecomment=[l]//,
  morecomment=[s]{/**}{*/}, % JavaDoc style comments
  morestring=[b]',
  morestring=[b]"
}[keywords, comments, strings]

\lstalias[]{ES6}[ECMAScript2015]{JavaScript}

\definecolor{mediumgray}{rgb}{0.3, 0.4, 0.4}
\definecolor{mediumblue}{rgb}{0.0, 0.0, 0.8}
\definecolor{forestgreen}{rgb}{0.13, 0.55, 0.13}
\definecolor{darkviolet}{rgb}{0.58, 0.0, 0.83}
\definecolor{royalblue}{rgb}{0.25, 0.41, 0.88}
\definecolor{crimson}{rgb}{0.86, 0.8, 0.24}

\lstdefinestyle{JSES6Base}{
basicstyle=\small,
  backgroundcolor=\color{white},
  basicstyle=\ttfamily,
  breakatwhitespace=false,
  breaklines=true,
  captionpos=b,
  columns=fullflexible,
  commentstyle=\color{mediumgray}\upshape,
  emph={},
  emphstyle=\color{crimson},
  extendedchars=true,  % requires inputenc
  fontadjust=true,
  frame=single,
  identifierstyle=\color{black},
  keepspaces=true,
  keywordstyle=\color{mediumblue},
  keywordstyle={[2]\color{darkviolet}},
  keywordstyle={[3]\color{royalblue}},
  numbers=left,
  numbersep=5pt,
  numberstyle=\tiny\color{black},
  rulecolor=\color{black},
  showlines=true,
  showspaces=false,
  showstringspaces=false,
  showtabs=false,
  stringstyle=\color{forestgreen},
  tabsize=2,
  title=\lstname,
  upquote=true  % requires textcomp
}

\lstdefinestyle{JavaScript}{
  language=JavaScript,
  style=JSES6Base
}
\lstdefinestyle{ES6}{
  language=ES6,
  style=JSES6Base
}

\usepackage{color}
\usepackage{algorithm,algpseudocode} %% select one of them

\renewcommand\footnotetextcopyrightpermission[1]{} % removes footnote with conference information in first column
\usepackage{hyperref}
\hypersetup{
     colorlinks = true,
     linkcolor = blue,
     anchorcolor = red,
     citecolor = magenta,
     filecolor = blue,
     urlcolor = black,
}

\begin{document}

\title{Towards Web3 Applications: Easing the Access and Transition}

%=================================================
%author
%=================================================

%\begin{comment}  % priority

\author{Guangsheng Yu}
\orcid{0000-0002-6111-1607}
\affiliation{%
  \institution{Data61, CSIRO}
  \city{Sydney}
  \state{NSW}
  \country{Australia}
  \postcode{2121}
}
\email{Saber.Yu@data61.csiro.au}

\author{Xu Wang}
\affiliation{%
  \institution{GBDTC, University of Technology Sydney}
  \city{Sydney}
  \state{NSW}
  \country{Australia}
  \postcode{2007}
}
\email{Xu.Wang-1@uts.edu.au}

\author{Qin Wang}
\affiliation{%
  \institution{Data61, CSIRO}
  \city{Sydney}
  \state{NSW}
  \country{Australia}
  \postcode{2121}
}
\email{qinwangtech@gmail.com}

\author{Tingting Bi}
\affiliation{%
  \institution{Data61, CSIRO}
  \city{Melbourne}
  \state{VIC}
  \country{Australia}
  \postcode{3168}
}
\email{Tingting.Bi@data61.csiro.au}

\author{Yifei Dong}
\affiliation{%
  \institution{GBDTC, University of Technology Sydney}
  \city{Sydney}
  \state{NSW}
  \country{Australia}
  \postcode{2007}
}
\email{Yifei.Dong@uts.edu.au}

\author{Ren Ping Liu}
\affiliation{%
  \institution{GBDTC, University of Technology Sydney}
  \city{Sydney}
  \state{NSW}
  \country{Australia}
  \postcode{2007}
}
\email{RenPing.Liu@uts.edu.au}

\author{Nektarios Georgalas}
\affiliation{%
  \institution{Applied Research, British Telecom}
  \city{Martlesham}
  \state{Woodbridge}
  \country{UK}
  \postcode{IP5 3RE}
}
\email{nektarios.georgalas@bt.com}

\author{Andrew Reeves}
\affiliation{%
  \institution{Applied Research, British Telecom}
  \city{Martlesham}
  \state{Woodbridge}
  \country{UK}
  \postcode{IP5 3RE}
}
\email{andrew.reeves@bt.com}

%\end{comment}

%=================================================
%abstract
%=================================================

\begin{abstract}
Web3 is leading a wave of the next generation of web services that even many Web2 applications are keen to ride. However, the lack of Web3 background for Web2 developers hinders easy and effective access and transition. On the other hand, Web3 applications desire encouragement and advertisement from conventional Web2 companies and projects due to their low market shares. In this paper, we propose a seamless transition framework that transits Web2 to Web3, named \textsc{WebttCom}\footnote{\textsc{WebttCom} stands for \textbf{Web}2 (\textbf{T}wo)--Web3 (\textbf{T}hree) \textbf{Com}municator.}, after exploring the connotation of Web3 and the key differences between Web2 and Web3 applications. We also provide a full-stack implementation as a use case to support the proposed framework, followed by performance evaluation and surveys with $\sim$1,000 participants that show $\sim$80\% positive and $\sim$20\% neutral responses. We confirm that the proposed framework \textsc{WebttCom} addresses the defined research question, and the implementation well satisfies the framework \textsc{WebttCom} in terms of strong \textit{necessity}, \textit{usability}, and \textit{completeness} based on the survey results.
\end{abstract}




%=================================================
\maketitle
%=================================================   

%------------------------------
\input{sections/Introduction}

\input{sections/Preliminary}

\input{sections/Design}

\input{sections/UseCase}
\input{sections/Discussion}
\input{sections/RelatedWork}
\input{sections/Conclusion}

%================================================
%=================================================
% \newpage
\bibliographystyle{unsrt}
\bibliography{bib.bib}

%================================================
% \newpage
\vspace{2em}
\section*{Appendix}
\input{sections/Appendix-1}
\input{sections/Appendix-2}

\end{document}

%% file: sections/Introduction.tex
\section{Introduction}\label{sec:intro}

%{\color{blue} (This section should introduce blockchain, Web2/web3, challenges to transit Web2 to web3 or connect Web2 and web3, motivation to come up with this design pattern or architecture, and contributions.)}

Web3 has drawn intensive attention from communities and investors. As an umbrella term, Web3 covers a series of blockchain-based decentralized applications (DApps), services, and economics \cite{weyl2022decentralized,wang2019survey,wang2022exploring,bc-abe,bc-lora} that bring significant impacts on both traditional finance and cryptocurrency markets. To date (as of Mar 2023), over $12,143$ DApps\footnote{Data source [Mar 2023]: DappRadar \url{https://dappradar.com/industry-overview}.} have been developed on-chain and $285,152$ smart contracts are deployed across $48$ protocols. A total of $1.67$M users are actively interacting with the smart contracts within 24 hours, as evidenced by their wallet addresses. In this sense, Web3 impresses users by providing such a \textit{connect the wallet} button on the upper-right corner of each webpage. Users can use DApps through embedded wallet entries by invoking specific functions that are deployed on blockchain-engined platforms (e.g., Ethereum \cite{wood2014ethereum}). The shift of backend servers from centralized clouds to decentralized chains has mostly distinguished Web3 and previous web styles.

Following its narrative connotation, we observe that Web3 users still occupy a pretty small percentage (appr. 0.03\% of $4.95$B\footnote{Data source: Digital 2022: Global Overview Report \url{https://datareportal.com/reports/digital-2022-global-overview-report}.} Internet users) over the Internet. The constraints majorly come from its incompatibility: (i) a typical Web3 application cannot be smoothly applied in a traditional web context due to the absence of blockchain engines; conversely, (ii) a traditional Web2 application can hardly be integrated with blockchain due to the lack of proper Application Programming Interfaces (APIs). Rational developers would start their work on wide-adoption applications, namely Web2 Apps, for higher user exposure and more potential revenues rather than sparing efforts on Web3 applications that are uncertain. Such concerns motivate this work: 

\smallskip
\textit{How to ease the usage of Web3 applications and achieve a smooth transition of applications between the Web2 and Web3 space? }
\smallskip

We investigate the bottlenecks of the transition between Web2 and Web3 by diving into their distinctions in design and implementation. For most Web2 applications, user management, data manipulation, and private-preserving policies are constructed upon centralized databases in a closed manner. In contrast, Web3 DApps usually apply asymmetric-encryption-based identity to establish more secure and robust user management in a decentralized manner among many parties without prior trustworthiness~\cite{10092670,10.1145/3551902.3564802,LIU2022102090, liu2022bgra}. Data manipulation differs from that of Web2 applications due to the immutability of data storage in Web3, and access control also requires new approaches to partition the visibility. 
In addition, existing Web3 DApps lack flexible mechanisms to smooth the workflow of documenting Restful APIs and test suites during the development phase. 
Therefore, the incompatible user management strategies, execution procedures for data manipulation, private-preserving policies, and the lack of Web3-compatible API tools have greatly retarded the smooth transition of applications in different domains.

To fill the above gaps, in this work, we propose a practical solution to integrate both Web3 and Web2 applications and related services seamlessly. We deconstruct the Web2 architecture and extract three major components: frontend APIs, backend servers, and supplementary databases. Aligning with the core principles of Web3, we accordingly modify these components to enable seamless integration with blockchain engines. Specifically, we design three types of adjustable components: a SaaS module for integrating Web2 software by providing generalized APIs, a backend interpreter that interprets and forwards requests from Web2 to Web3, a blockchain layer that configures self-governed policies via smart contracts as well as computes on-chain calculations through chain Software Development Kits (SDKs). Our proposed solutions can greatly promote the transition from classic Web2 applications to the Web3 space without redundant development or complicated middleware. In short, we highlight the \textit{\textbf{contributions}} as follows.

\begin{itemize}

\item \textit{We explore the connotation of Web3 by investigating plenty of in-the-wild Web3 projects.} As a new concept, we compare existing Web2 solutions and so-claimed Web3 projects to determine the root features of Web3 applications and their dependencies, extracting their differences from classic Web2 applications.  

\item \textit{Based on comprehensive investigation, we propose a seamless transition framework that transits Web2 to Web3}, named \textsc{WebttCom}. Our proposed solution establishes an interpreter to bridge the Web2 applications and the Web3 backend engines. In particular, the proposed framework \textsc{WebttCom} can offer effective and reliable access control and user management across the decentralized Web3 and centralized Web2 and provide an approach to conduct transition with existing popular SaaS and the framework. While operating, \textsc{WebttCom} can also effectively improve production by automatically generating API documents for developers and communities. 

\item \textit{We provide a full-stack implementation ranging from front and backend APIs, structure design, and smart contract programming to the on-cloud dockerized deployment}. Code size reaches up to 11,351 lines\footnote{Specifically, we present the detailed size distribution. The front end takes 4 MB with image resources,  while 253 KB with code only (5367 lines). The backend is a 1.3 MB project and 249 KB for codes (4831 lines). Web3 takes 265KB (1153 lines). Note that the source code is set confidential due to the non-disclosure agreement with British Telecom (BT).}. The system is applied to daily service management processes and promotes the establishment of effective, flexible, reliable, and trustworthy Web3-driven applications. Our prototype has been inspected by British Telecom (BT) in practice.

\item \textit{We further conducted a quantitative performance evaluation and a qualitative evaluation from the perspectives of managers, investors, and developers through surveys targeting 1,000 practitioners.} The survey feedback shows that the research question is well satisfied by the proposed framework \textsc{WebttCom}, and the presented full-stack implementation proves its strong capability of smoothing the transition in terms of \textit{necessity}, \textit{usability}, and \textit{completeness}.

%{\color{blue} Contribution 1: transition (we provided a framework that XXX transits Web2 to 3, and API documentation); contribution (RQ1 and RQ2 ATM)}

%\item {\color{blue} Contribution 2: validate an industrial case with respect to the framework, The case implements...and...and.... We also conduct an evaluation from developers' perspective. The survey feedback shows that...}

% \item \textit{We explore the connotation of Web3 by investigating plenty of in the wild Web3 projects.} 

% \item \textit{We summarise major design patterns and characteristics in building Web3 applications.} In particular, we focus on how to build a Web3 application that can seamlessly connect to existing Web2 infrastructure. We highlight several key properties during the development as well as give a catalog of empirically-justified design patterns for building Web3 applications.

% \item \textit{We implement a prototype system as the solid use case.} Specifically, we design a decentralized product inventory system, bridging the gap between decentralized trustworthiness and traditional Web2 applications.

% \item  \textit{We further discuss notices for implementations and point out related limitations for new Web3 patterns.} 
% \item  Productivity improvement: the proposed approach can effectively improve the production of software development and documentation, in terms of API XXX documents automatically generation. 
\end{itemize}

The remainder of this paper is organized as follows. Section~\ref{sec:premili} provides the Web3 basics. Section~\ref{sec:design} proposes the research methodology and presents our research question. Section~\ref{sec:usecase} introduces a use case that implements the new pattern, followed by discussions about the implementation notices and limitations shown in Section~\ref{sec:discussion}. Section~\ref{sec:related} provides existing studies related to this work. Section~\ref{sec:conclusion} concludes this paper and highlights our contributions.

%% file: sections/Preliminary.tex
\section{Approaching Web3: A PRELIMINARY}\label{sec:premili}

In this section, we show the Web3 basics by comparing with Web2 and providing a typical Web3 instance. 

\subsection{Differences between Web 1/2 and Web3}

Traditional Internet, including so-claimed Web1 and Web2, has been developed for decades. Web1 is regarded as a suite of \textit{read-only} protocols that contain static sites to present images, text, and videos. Users search for the targets by accessing web portals. Web2 changes the way of interaction by enabling user-generation content (UGC). Users can publish their original content, such as images, reviews, testimonials, or even podcasts, on social media websites (e.g., Facebook, Twitter). In this sense, Web2 is regarded as \textit{read-write}. Web3 differs from previous styles by adding features of \textit{ownership} and \textit{transfer}. Users will create self-controlled accounts, generally in the forms of \textit{wallets}, to manage digital assets and virtual data. Rather than relying on centralized servers, Web3 users can freely transfer their assets under the governance of smart contracts, which brings the advantages of auto-execution, being accountable, and being globally verified. These smart contracts connect both upper-layer DApps and underlying blockchain platforms.

\begin{table}[!hbt]
    \vspace{-0.2cm}
    \centering
    \caption{Comparisons among Web1/Web2/Web3 and Our Work}\label{tab-comparison}
    \resizebox{0.8\linewidth}{!}{
    \begin{tabular}{lccr} 
    \toprule
    \multicolumn{1}{c}{\quad\textbf{}\quad}   & \multicolumn{1}{l}{\quad\textbf{Functions}\quad} & \multicolumn{1}{l}{\quad\textbf{Architecture}\quad} & \multicolumn{1}{l}{\quad\textbf{Instance}\quad}  \\
    \midrule
   \textbf{\textit{Web1}} & read &  client-server & Yahoo    \\
   \textbf{\textit{Web2}} & read/write & client-server & Facebook, Google\\
   \textbf{\textit{Web3}} & read/write/own/transfer & client-SC-chain  & Ethereum, BSC    \\
   \textbf{\textit{Web2$\to$3}} & read/write/own/transfer & client-Int.-SC-chain  &  \textbf{\textsc{WebttCom}}    \\
   
   \bottomrule
    \end{tabular}
    }
    \vspace{-0.2cm}
\end{table}

\subsection{Typical Web3 Architecture}

Tradition web architecture is based on a client-server model. The client is used for sending and receiving requests, while the server is used to process these requests and corresponding logic. The server side, also known as the backend, covers many fundamental aspects like operating systems (Windows, Linux), platforms (.Net, LAMP), and storage. API is to connect the application tier to servers. In contrast, the Web3 architecture replaces centralized backend servers with distributed ledgers. The backend contains two sectors, smart contracts (SC) for defining logic and rules, and blockchain platforms for processing transactions and achieving consensus. Web3 is more complex than traditional Web2 due to its complete decentralization, which requires dealing with the consistency problem~\cite{WANG2021112}. In this work, we aim to ease the transition between the Web2 application tier and blockchain-backend systems. We establish an interpreter (short for Int.) to connect them seamlessly.

\subsection{Essential Component}

We provide basic building blocks to construct Web3 applications: a series of blockchain-specific components covering blockchain, smart contracts, on-chain applications, and clients. 

\smallskip
\noindent\textbf{Blockchain.} Blockchain is a digital ledger that operates in a decentralized manner to securely and transparently record transactions \cite{bonneau2015sok}. Transactions are recorded as blocks, which are subsequently organized into a hierarchical structure. The blocks are arranged in a chronological and unalterable sequence to form the blockchain. To add a new block to the chain, a process called "mining" is used, which involves solving complex mathematical problems to validate transactions and add new blocks to the blockchain. This mining process is regulated by a consensus mechanism, which sets rules to ensure that all participants in the blockchain network agree on the validity of the transactions and their order in the blockchain.

\smallskip
\noindent\textbf{Smart contract.} Smart contracts are computer programs that automatically enforce the terms of an agreement between parties. They are used to speed up, verify, or execute digital negotiations. Ethereum developed smart contracts on the blockchain system by using Turing-complete scripting languages to achieve complex functionalities and execute thorough state transitions via consensus algorithms, resulting in final consistency. Smart contracts enable unfamiliar parties to conduct fair exchanges without the need for a trusted third party. They have a broad range of applications, including financial services \cite{werner2021sok}\cite{jiang2023decentralized}, security protocols \cite{li2022smart}, and decentralized governance \cite{kiayias2022sok}. Smart contracts are viewed as a disruptive technology that could revolutionize many industries by improving efficiency, reducing costs, and eliminating the need for intermediaries.

\smallskip
\noindent\textbf{On-chain application.} On-chain applications, also known as DApps, are applications that run on blockchain systems \cite{mohanta2019blockchain}. Unlike traditional applications, which are centralized and controlled by a single entity, on-chain applications are decentralized and operate on distributed networks. The applications are designed to be transparent, secure, and trustless. They use smart contracts to execute code and transactions, and the consensus mechanism of the underlying blockchain network to verify and validate these transactions. Thanking the nature of decentralization, they are resistant to censorship, tampering, and other forms of malicious activity~\cite{2023arXiv230608056K}. For specific usage, on-chain applications include various types such as NFT \cite{wang2021non}, gaming, social networking, etc.

\smallskip
\noindent\textbf{Light client.}  A light client \cite{chatzigiannis2021sok} plays a crucial role in the world of Web3 and blockchain technology, acting as an intermediary between user requests and the back-end servers or blockchains. Its primary function is to forward user requests to their appropriate destination without engaging in any logic processes. Generally, a light client is represented by a wallet \cite{karantias2020sok} in the context of Web3 or blockchain. It is supported by locally running light nodes, which synchronize information with full nodes. This makes using a light client highly advantageous in resource-constrained environments, such as different hardware devices, as it can reduce the costs of performing complex computations on-chain~\cite{WANG2019100109,10.1007/978-3-030-36938-5_47,9526860}.
In the Web3 ecosystem, it is typical for users to connect their wallets to perform interactive actions on the website. By understanding the role of a light client, users can navigate the complexities of Web3 and blockchain technology with ease.

%% file: sections/Design.tex
\section{Research Design}
\label{sec:design}
In this section, we conduct an exploratory study \cite{easterbrook2008selecting}, which proposes a new Web3 driven framework, named \textsc{WebttCom}, for implementing applications; and we define the research question (RQ) for identifying and extracting the evidence (e.g., the benefits of Web3 applications) for evaluating our proposed Web3 framework.

\subsection{Research Question}
\label{subsec_research_design_RQs}

This study aims to design and analyze whether our Web3 framework is effective and practical regarding quality attributes (i.e., smooth transition) and development productivity. Such characterization of Web3 will shed light on future Web3 applications and developments. Specifically, this work aims to address the Research Question (RQ) as:

\smallskip
\noindent\textbf{\textit{Given the proposed Web3 framework (i.e., \textsc{WebttCom}), is it effective, regarding the transition of Web2 to Web3 application development?}}

\begin{itemize}
\item \textbf{Web2 to Web3 transition}: Given that many developers do not have relevant development background for Web3 applications, is the proposed framework practical and helpful for them to transit the Web2 to Web3 applications for meeting and implementing the specific requirements?

\item \textbf{Data privacy and governance}: In the evolution from Web2 to Web3, various issues related to data privacy and security, data privacy and governance in Web3 is one of the most discussed challenges for different stakeholders. As such, to address this concern, our RQ can explore whether the proposed Web3 framework (i.e., \textsc{WebttCom}) ensures data privacy and governance.

\item \textbf{Development productivity}: If the proposed Web3 framework impacts developers' productivity? For example, (1) practitioners' daily development tasks and documentation. (2) problem-solving for Web3 application development. 
\end{itemize}

\subsection{Study Design Process}

\begin{figure}[!hbt]
	\centering
	\includegraphics[width=0.9\linewidth]{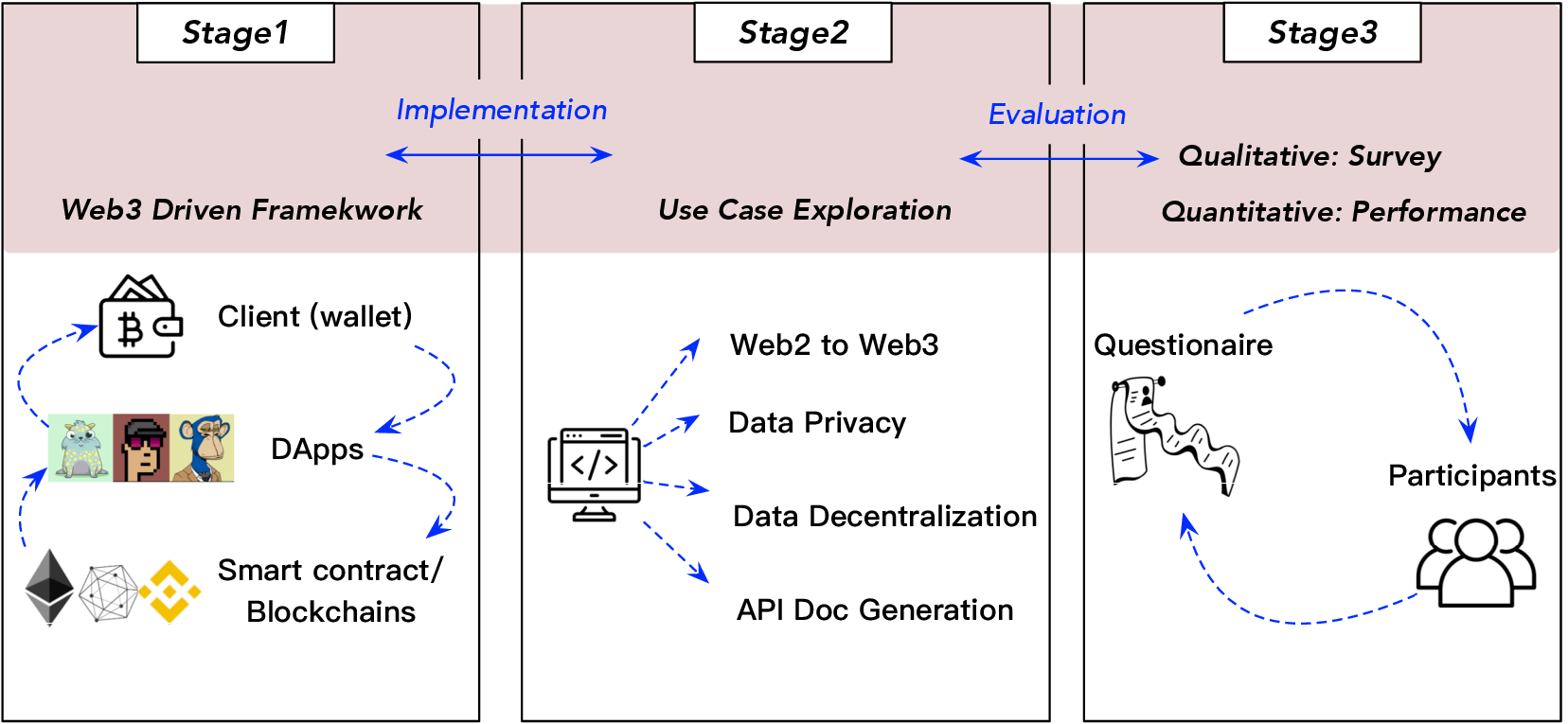}
	\caption{Overview of the study design}
	\label{fig:overview_of_researchdesign}
\end{figure}

Our research methodology consists of three stages, as depicted in Fig. \ref{fig:overview_of_researchdesign}. In the first stage, we proposed a Web3 transition framework (i.e., \textsc{WebttCom}), which includes an interpreter to help practitioners transit and bridge the Web2-based to Web3-based applications. We implemented a full-stack project based on the proposed \textsc{WebttCom} in the second stage. In the last stage, to evaluate the effectiveness of the proposed \textsc{WebttCom} and the feasibility of the full-stack project, we conducted a quantitative performance evaluation and carried out a qualitative survey with 1,000 practitioners, including Venture Capitals (VCs), business managers, and developers, to gather their feedback and opinions.

\smallskip
\noindent\hangindent 1em \textbf{Phase 1: Web3 driven framework.}
We proposed a Web3 framework (\textsc{WebttCom}), which allows developers to implement effective and reliable applications. The framework guarantees:

\begin{itemize}
\item Web2 to Web3 smooth transition: our framework provides a trustworthy transition. 
\item Blockchain backend: the framework ensures the decentralization of  Web3-driven applications.
\end{itemize}

The details of the framework are described in Section~\ref{sec_framework}.

%\smallskip
\noindent\hangindent 1em \textbf{Phase 2: Web3 application implementation.}
In this phase, we implemented a full-stack application, which is based on the Web3 framework that we proposed. The details of the application are described in Section~\ref{sec_implementation}.

%\smallskip
\noindent\hangindent 1em \textbf{Phase 3: Performance evaluation and surveys.}
As depicted in Fig. \ref{fig:overview_of_researchdesign}, in this phase, we gave the overall quantitative performance evaluation. At the same time, we invited 1,000 practitioners to participate in qualitative surveys to confirm the effectiveness of our proposed Web3-driven framework and implementation.

\begin{itemize}
\item We invited 1,000 practitioners from BT, universities, and VC members in Australia and China, as well as developers from GitHub and those on the mailing list, to participate in our surveys. These surveys were anonymous and the survey results were based on the valid respondents.
\item We designed a questionnaire consisting of three main questions (see Section~\ref{sec_questions}) to get developers' opinions on the effectiveness of our proposed framework and the implementation we developed.
\item We applied qualitative data analysis and consistent comparison to summarize the statements from the survey result regarding the productivity of using our framework and application.
\end{itemize}

The questionnaire consists of three parts of questions: 
\begin{itemize}
\item \textit{Part 1:} We asked demographic questions, such as the surveyees’ experience in Web3 development and management.
\item \textit{Part 2:} We asked open-ended questions to understand their opinions on Web3 design and development in practice. 
\item \textit{Part 3:} We prepared candidate topics by carefully reading the contents of representative textbooks. We picked a list of topics not explicitly mentioned in the open discussion and asked the participants to discuss those topics further. At the end of each survey, we thanked the participants and briefly informed them what we planned to do with his/her response.
\end{itemize}

%% file: sections/UseCase.tex
\section{Research Result}\label{sec:usecase}

In this section, 
we propose a new framework, named \textsc{WebttCom}, which not only smooths the transition between Web2 and Web3 but also achieves high data privacy and data governance and improves development productivity. A full-stack implementation is presented as a use case to support the framework. We also conducted surveys where the intended 1,000 practitioners were invited to confirm the effectiveness of the proposed framework and implementation.

\subsection{A New Framework: \textsc{WebttCom}}
\label{sec_framework}

\begin{figure*}[t]
	\centering
	\includegraphics[width=\linewidth]{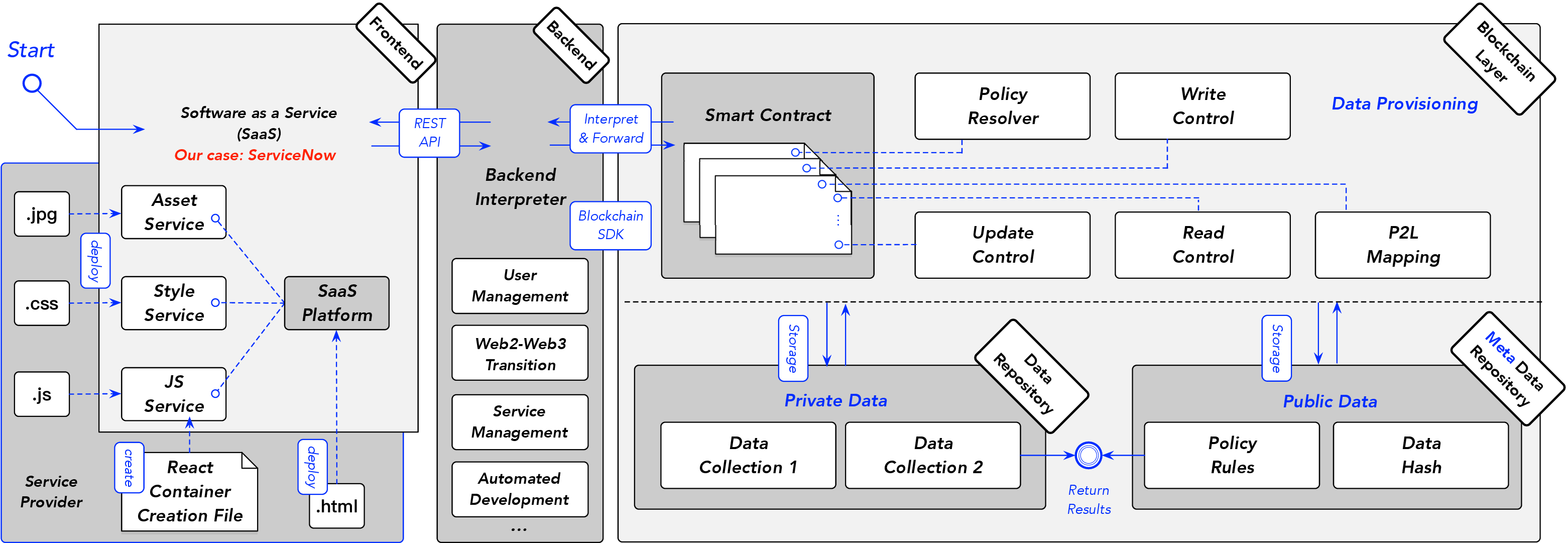}
	\caption{Overview of the instantiated service management platform.}
	\label{fig:btoverview}
\end{figure*}

A new framework, named \textsc{WebttCom} (Web2 to Web3 Communicator), is proposed as a result. \textsc{WebttCom} provides trusted and reliable service management across multiple organizations in a decentralized manner and promotes the transition from the existing Web2 form to Web3. Note that ``transition'' does not mean the existing Web2 applications must be completely reconstructed. Instead, a new interpreter is attached in the middle to offer reliable communication, smoothly integrating Web2 applications to Web3 platforms and transiting Web2 applications that would have been slowly drifting apart from decentralization into a Web3-empowered structure. In particular, the structure of \textsc{WebttCom} consists of a blockchain layer, a backend interpreter, and a Web2 Software as a Service (SaaS), as in Fig.\ref{fig:btoverview}.  

The blockchain layer stores service management data and supportive data and controls access to the service data. 
To this end, the blockchain layer contains a smart contract for data provisioning, a private data repository for service data, and a public data repository for data access policies and data hash checksum. 
In the smart contract, the policy resolver module extracts attributes of requests, such as requested data, request location, and user affiliation, and identifies effective access control policies.
The write control module enforces data storage policies, such as data storage location and data expiration.
The update control module takes charge of data updates, while the read control module enforces data access policies.
The Physical-to-Logical (P2L) mapping module updates data presentation according to pre-defined physical data formats for data storage and logical data formats required by the SaaS. Note that the blockchain layer is designed to be blockchain-agnostic, allowing for seamless integration with any blockchain platform empowered by automated smart contracts and applicable functionalities such as user management, access controls, external integration, etc. This design ensures flexibility and adaptability, as the system can leverage the strengths of various blockchain platforms without being constrained by any single platform.
% In the data repositories, the private data repository collects all service data following access control policies and the metadata repository saves access control policies and data hash checksum.

% The backend acts as a Web2-Web3 interpreter forwarding requests from the Web2 SaaS to the Web3 smart contract and formatting Web3 results for Web2 SaaS, as in the Web2-Web3 transition module. The backend interpreter meanwhile manages Web2 and Web3 users for the transition.
% On the Web2 side, the backend interpreter provides REST API~\cite{wang2014developers} to the SaaS. The backend interpreter employs automated development technology for the automatic route and document generation.
% The backend interpreter communicates with the blockchain via the blockchain SDK.

The backend, functioning as a Web2-Web3 interpreter, bridges the gap between traditional Web2 SaaS platforms and the emerging Web3 ecosystem by facilitating seamless communication between the two. This is achieved through the Web2-Web3 transition module, which is responsible for forwarding requests originating from Web2 SaaS applications to Web3 smart contracts and subsequently formatting the resulting Web3 responses in a manner that can be easily understood by Web2 SaaS applications.
In order to effectively manage users during the transition phase, the backend interpreter maintains separate user management systems for both Web2 and Web3 users, with a completed synchronization at the end. This ensures that each user category's unique attributes and requirements are adequately addressed while preserving the overall user experience.
For the Web2 side, the backend interpreter offers a comprehensive REST API for SaaS applications to interact with. This API enables SaaS platforms to send requests and receive responses from the Web3 environment without directly interacting with Web3 technologies. This simplifies the integration process and allows developers to focus on their core application logic.
To further streamline the development process, the backend interpreter incorporates automated development technology that automatically generates routes and documentation. This significantly reduces the time and effort required by developers to create, maintain, and update their APIs, allowing them to concentrate on delivering high-quality services to their users.
When it comes to communicating with blockchain networks, the backend interpreter leverages blockchain Software Development Kits (SDKs). These SDKs provide a set of pre-built tools and libraries that make it easier for developers to interact with various blockchain protocols and smart contracts. By utilizing these SDKs, the backend interpreter can efficiently query, submit transactions to, and retrieve data from the blockchain while maintaining high security and reliability.
In other words, the backend interpreter plays a crucial role in bridging the divide between Web2 SaaS applications and the emerging Web3 ecosystem. By offering a seamless communication channel between the two environments, managing user transitions effectively, and providing robust development tools, the backend interpreter enables the integration of Web3 technologies into existing Web2 applications, paving the way for a decentralized and interconnected digital landscape.

Web2 SaaS platforms offer sophisticated service management interfaces that allow users to easily submit, view, and process service tickets after authenticating themselves on the platform. These interfaces are designed to provide a seamless user experience, making it simple for users to access and manage their service requests.
To ensure adaptability and compatibility with the emerging Web3 technologies, the service management functionality is implemented as a modular, plug-and-play component. This design approach enables the service management module to be easily integrated with popular service management platforms, such as ServiceNow React Container~\cite{servicenow}, while allowing for incorporating Web3-enhanced features.

\subsection{Implementing \textsc{WebttCom}: A Use Case in a Service Management System}
\label{sec_implementation}

\begin{figure*}[t]
	\centering
	\includegraphics[width=1\textwidth]{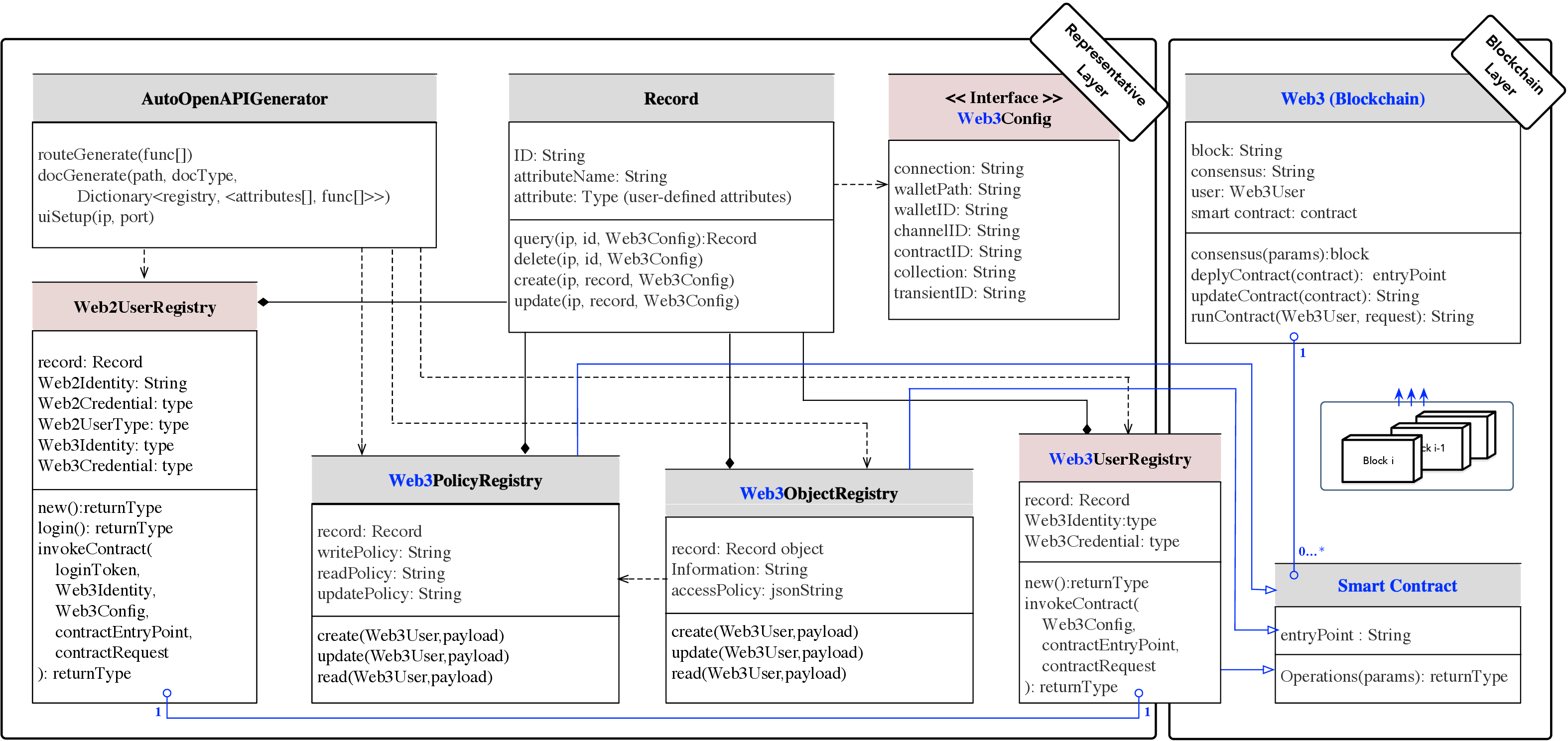}
	\caption{The class diagram of the instantiated service management platform.}
	\label{fig:class-diagram}
\end{figure*}

An implementation of the proposed framework \textsc{WebttCom} is discussed in which a trusted and reliable service management platform across BT Global, BT Australasia, and BT Australasia customers is demonstrated. 
Managing the services (e.g., inventory/ordering/ticketing) that BT deploys is challenging due to a lack of trust in the data to execute these processes. Conflicting information, lack of real-time parameters of the state of the various devices, and complex disputes between two or more parties are just a few ingredients of these challenges. BT’s services can consist of equipment from different manufacturers deployed in different countries, managed by different secondary service providers, operating under different SLAs, and serving different customers. Correct management of these complex services has proven to be one of the key challenges for BT’s engineering teams and is one of the key factors for the company’s international reputation and success. Currently, these services are managed by legacy and non-integrated systems that often cannot operate in real-time to meet global customers’ demands.

The HyperLedger Fabric~\cite{androulaki2018hyperledger} is set up with one Fabric ordering service\footnote{The ordering service validates transactions and assembles valid transactions into ordered blocks.}, and organizations representing BT Global, BT Australasia, and BT Australasia customers in the same channel\footnote{A Fabric channel is a private sub-network for permitted members.} for the Web3 service. Each ordering service and organization has one Certificate Authority (CA) node and one peer node. Service ticket data are recorded in the Hyperledger Fabric in real time and are certified by the data hash. With Web3 technology, all organizations have a consistent view of the service data. The platform is developed in TypeScript 4.4 using Visual Studio Code 1.69.1 and runs on the elastic servers of Amazon Web Service (AWS). MySQL 5.7 is selected to handle the Web2-based registries, whereas the Web3-based registries rely on the native CouchDB of HyperLedger Fabric 1.4.

The implementation design of a service management system, as a use case of the proposed \textsc{WebttCom}, is illustrated in Fig.\ref{fig:class-diagram} using the class diagram. \textit{Blockchain} maintains a blockchain network with several registered \textit{SmartContract}.
% , e.g., HyperLedger Fabric in the use case. 
\textit{SmartContract} is inherited by three classes, i.e., \textit{Web3UserRegistry}, \textit{Web3ObjectRegistry}, and \textit{Web3PolicyRegistry}. The \textit{Web3UserRegistry} registers the user Web3 information in Hyperledger Fabric CAs and interacts with the corresponding customizable off-chain-stored \textit{Web2UserRegistry} in which the local user management is conducted. All objects, including the product inventory, product order, service inventory, and trouble ticket service, are on-chain-stored in \textit{Web3ObjectRegistry} with dedicated policies of access control stored in \textit{Web3PolicyRegistry}. All registries comprise \textit{Record} in which the user-defined data record attributes are stored. Accessing the attribute values in \textit{Record} relies on \textit{Web3Config} in which the settings of \textit{Blockchain} are defined. \textit{AutoOpenAPIGenerator} auto-generates the OpenAPI standard~\cite{openapi} documentation for modules including \textit{Web2UserRegistry}, \textit{Web3UserRegistry}, \textit{Web3ObjectRegistry}, and \textit{Web3PolicyRegistry}.

% Web3 layer

% Interpreter

% Web2 layer

\subsubsection{Attribute-based Access Control in Web3}
% \sy{This paragraph may need to be moved to elsewhere as it only introduces the Web3 background. } 
In the Hyperledger Fabric, each organization has one private data collection~\cite{fabricpd}, which excludes other organizations, such that private business data are only saved and managed by the permitted organization.  
All blockchain members can access public on-chain data.
Web3 users are registered to CAs with their attributes, including organizations, departments, and user types. The attributes are used for the attribute-based access control to the service data. Each Web3 user has a unique private key to sign Web3 transactions and access Web3 services.

\textit{Smart Contract: }
A service smart contract is developed to manage service tickets. The exposed entry points of the smart contract are shown in Listing~\ref{code_contractentry}. The smart contract extends the Fabric \emph{Contract} class and exports functions as APIs. 

In the smart contract, create/update/read functions are developed for \emph{Web3PolicyRegistry} and \emph{Web3ObjectRegistry}, as shown in Fig.\ref{fig:class-diagram}. 
% As shown in Fig.~\ref{}, \emph{Web 3PolicyRegistry} collects access control policies, including write policy, read policy and update policy, while \emph{Web 3PolicyRegistry} saves service data and accompanying access policies. 
% Both registries support create, update and read operations. 
\emph{Web3PolicyRegistry} is implemented with the Fabric public data scheme where all blockchain peers have a copy of the public data, while \emph{Web3ObjectRegistry} is realized with the Fabric private data scheme where only authorized peers have data copies.
Variable \emph{ctx} collects the context of the transaction calling the smart contract, including user attributes and timestamp for access control and transient service data\footnote{Transient data is a type of data for Fabric private data collection, which can be read by smart contracts but does not be recorded in on-chain transactions.}. 
Variable \emph{attributes} provides Web2 attributes for access control, such as geolocation information. Variable \emph{publicPayload} gives the data requested by smart contracts and recorded on the Fabric blockchain, such as access control policies.

% Create/Update/Read functions polices of access control \textit{Web3PolicyRegistry}, and business objects \textit{Web3ObjectRegistry} can be flexibly selected by a Web3-based controller via inversion of control, as shown in the return statements of List~\ref{code_contractentry}.

\textit{Access Control: }
The smart contract enables attribute-based access control (ABAC), where access control policies are designed based on user and environment attributes and organized as the \textit{Web3PolicyRegistry}.
A policy is either \textit{writePolicy}, \textit{readPolicy}, or \textit{updatePolicy}, and each of them has unique fields describing access control requirements. All the policy types also have general metadata, including an updated time, an effective time, and a priority level, for policy conflict resolver. 
All policies are created by organization admins and saved in the public data repository such that policies can be accessed by all Web3 users and applied to all private data collections.

The policy resolving process is shown in Listing~\ref{code_readcheck}.  
The policy resolver first extracts the Web3 user attributes and environment attributes of the transaction, such as transaction timestamp, from the \emph{ctx} variable. Next, the policy resolver module retrieves all related policies according to the policy IDs embedded in the request data. Then, the policy resolver identifies effective policies by checking Web3 user roles, effective timestamps, policy priority, and geolocations. At last, the policy resolver returns effective policies for the request.

The write control, update control and read control modules then process the request according to the request data in the request transactions and effective policies. 
The request is rejected if no effective policy is returned from the policy resolver module.
For a data write/update request, the write control and update control modules first identify the right private data collection from effective \emph{writePolicy} and \emph{updatePolicy}. The modules also check request attributes against the requirements specified in the effective policies and terminate the request for failed checks. The modules then extract transactional data and transient data from the request transaction and create a Web3 object with object information and related access policy IDs. Then, the modules can write/update the Web3 object to the data collection. 
For a data read request, the read control module gets the private data collection from the request transaction and reads the requested data out following the effective \emph{readPolicy}. The read control module also implements the P2L mapping on the data keys following the mapping rules stated in effective \emph{readPolicy}.

\subsubsection{Web2-Web3 Transition}

Recall that ``transition'' does not lead to a completed reconstruction of existing Web2 applications. It is the interpreter that establishes the communication and smoothly makes the Web2 applications empowered by Web3. Each organization has an interpreter to convert Web2 requests to Web3 requests and forward requests to specific Web3 nodes according to request attributes for Web3 processing. The interpreter then monitors Web3 processing results and creates returns for Web2 requests.  

% specifically focuses on the management of the connection between \textit{Web2UserRegistry} and \textit{Web3UserRgistry} upon the \textit{Web3Config}.

% \begin{lstlisting}[float=*,style=JavaScript, caption={Interpreter.},label={code_interpreter},basicstyle=\ttfamily\scriptsize]
%     export async function Trans(
%       _fabric: ITransient, context?: Required<Pick<IUserManagement, "userID" | "userPWD">>, fnc: string, ...args: string[]
%     ): Promise<string> {
%       let promise: Promise<string> = new Promise(async (resolve, reject) => {
%         try {
%           const fabricConCof: FabricConfiguration = new FabricConfiguration(
%             _fabric.connection,
%             _fabric.walletPath,
%             _fabric.walletID,
%             _fabric.channelName,
%             _fabric.scName,
%             _fabric.listen
%           );
%           const [gateway, connectionProfile, connectionOptions] = await getFabricConfig(context, _fabric);
%           await gateway.connect(connectionProfile, connectionOptions);
%           let network = gateway.getNetwork(fabricConCof.channelName);
%           let contract = (await network).getContract(fabricConCof.scName);
%           const channel = (await network).getChannel();
%           let result: Buffer | any;
%           result = await contract.submitTransaction(fnc, ...args);
%           await gateway.disconnect();
%           if (result) { resolve(result.toString()); }
%           else { resolve("Invoking ${fnc} succeeds."); }
%         } catch (error) { reject(error); }
%       });
%       return promise;
%     }
% \end{lstlisting}

\textit{Web2 APIs: }
The interpreter provides REST APIs supporting create, read, and update operations in Web2 applications and implements the Bearer Authentication scheme~\cite{bearer}.

A Web2 request includes payloads describing create, read, and update operations on Web3 objects and policies. Besides that, the Web2 request contains a bearer token for authentication by the interpreter. The bearer token also allows the interpreter to identify the corresponding Web3 identity and private key for signing Web3 transactions. The Web2 request also gives details for the corresponding Web3 request, including the channel name, smart contract name, and targeted private data collection, such that the interpreter can create correct Web3 request transactions and send them to the right Web3 node for processing.

\textit{Web3 Interaction: }
 Interpreters manage user identities and interact with the service smart contract using the Hyperledger Fabric SDK~\cite{fabricsdk}, as shown in Listing~\ref{code_interpreter}. The channel and service smart contract is specified by variables \emph{channelName} and \emph{scName} in lines 10-11.  

For every Web3 request, the interpreter needs to create a request transaction and sign the transactions with a valid Web3 user identity. To this end, the interpreter saves the Web3 user identities of all Web2 users in the organization, i.e., \emph{walletPath} and \emph{walletID} in Listing~\ref{code_interpreter}, and maintains the mapping between Web2 and Web3 identities. The interpreter is an organization admin user who can register and revoke users.
During the user registration process, a user provides user information and credentials to the interpreter, like general Web2 user registration. The interpreter first creates a new user record on itself and then sends the registration information to the organizational CA node to create a Web3 identity. Then, the interpreter keeps the user's Web3 wallet, including a private key, and records the mapping between the Web2 identity and Web3 identity. 

The interpreter needs to connect specific Web3 nodes that have requested private data collections. To this end, the interpreter maintains connection profiles\footnote{A connection profile describes the Web3 node to be connected, such as ID and IP/port.} to Web3 nodes from different organizations and records the mapping between private collections and the hosting Web3 nodes. When the interpreter receives a Web2 request, the interpreter identifies the data collection related to the request. The interpreter forwards the request to any Web3 node if the requested data is public. If the request points to some private data collection, the interpreter first identifies the hosting Web3 node and then forwards the request to the Web3 node using pre-configured connection profiles.

% \subsection{\sy{Transition with SaaS and API generation}}

\subsubsection{Transition with SaaS: ServiceNow-based Web Application}

ServiceNow is a popular cloud-based workflow automation platform for enterprises by streamlining and automating routine work tasks. Companies often seek a simple solution for connecting their existing ServiceNow applications to Web3 systems. As a result, we chose ServiceNow as our front-end system.
We intend to solve the following problems with our scheme:
\begin{itemize}
\item ServiceNow's existing system could be transformed in an efficient manner.
\item Programmers' coding can be made easier by following a few simple steps.
\end{itemize}

To resolve these issues, we implement a ServiceNow container as shown in the left part of Fig.\ref{fig:btoverview}.  We develop a web application using React and embedded it in ServiceNow. By using REST APIs, the embedded application can smoothly communicate with the backend and the interpreter. 

The web application was developed using React, an open-source JavaScript library for creating user interfaces with a large fan community. Therefore, most programmers are familiar with it and can easily obtain assistance from other community members.

A modern website deployment method - Webpack - is used to deploy this web application. Instead of deploying all resource files directly to the web server, Webpack binds multiple files into a single deployment file. Multiple HTML files are combined into one HTML file, and multiple Javascript files are combined into one Javascript file. This deployment method simplifies the resource access path, which is exactly what we require.
\begin{itemize}
\item The deployment utilizes two ServiceNow concepts: the UI page and the web service.

\item The ServiceNow UI page allows developers to customize web pages using HTML or XML. In this case, we create one UI Page and insert the combined HTML code using Webpack.

\end{itemize}

The ServiceNow web services provide access to resources through REST APIs. Three web services have been developed: the Asset Service, the Style Service, and the JS Service. All images are accessible through the asset service. The styling service provides all CSS access. The JS service provides access to JavaScript code.

The web application has three functions: user login, policy management, and ticket management. As part of implementing these functions, the embedded application communicates with the interpreter using HTTP REST APIs.

\subsubsection{Backend and Automated Development Tools}
The Express framework offers a flexible Node.js web application with MySQL being used for \textit{Web2UserRegistry} and \textit{Web3Config}. Sequelize is used to provide effective connections to the Web2 local registry in the manner of an Object-relational Mapper (ORM). The platform is docker-based by default, and docker-compose offers an orchestration service for distributed and flexible docker containers.

The system also offers automated generation and documentation of OpenAPI-compliant REST APIs~\cite{openapi} for \textit{Web2UserRegistry}, \textit{Web3UserRegistry}, \textit{Web3ObjectRegistry}, and \textit{Web3PolicyRegistry} via
TSOA 3.11~\cite{tsoa}.
The TSOA framework integrates the OpenAPI compiler to construct Node.js server-side applications in type-safe TypeScript at runtime, as TypeScript is used for programming smart contracts at the Web3 side and backend/frontend at the Web2 side.
The function, \textit{routeGenerate}, can easily generate the API routes automatically with no pain based on the definition of controllers of all registries above; see the lower half of each registry in Fig.\ref{fig:class-diagram}. At the same time, the YAML-based OpenAPI documentation and test suite are also generated to ease API testing and smooth the development of transition via \textit{docGenerate}.

\subsection{Performance at a Glance}
The developed \textsc{WebttCom} and Hyperledger Fabric 1.4.6 have been deployed on a Mac mini, equipped with a 3.2 GHz 6-Core Intel Core i7 and 32 GB 2667 MHz DDR4 memory, utilizing Docker. \textsc{WebttCom} operates with Fabric SDK fabric-network version 1.4.0, while the Hyperledger Fabric connection profiles and the wallets of associated users are integrated into \textsc{WebttCom}. Within Hyperledger Fabric, the \emph{BatchTimeout} is configured to 2 seconds, establishing the maximum duration a blockchain node should wait for transactions to be mined into a block. Consequently, write operations can be confirmed in a maximum of 2 seconds.

\begin{table}[!hbt]
    \centering
    \caption{APIs in the Performance Evaluation}
    \label{tab-apis}
\begin{tabular}{|c|c|c|c|c|}
\hline
\textbf{API} & \textbf{R/W} & \textbf{Public/Private} & \textbf{Payload} & \textbf{Size} \\ \hline
Get a policy      & Read                & Public                  & 1 policy         & 1KB                \\ \hline
Get all policies  & Read                & Public                  & 91 policies      & 56.97KB            \\ \hline
Get all tickets   & Read                & Private                 & 8 tickets        & 5.67KB             \\ \hline
Create a policy   & Write               & Public                  & 1 policy         & 1KB                \\ \hline
Create a ticket   & Write               & Private                 & 1 ticket         & 2KB                \\ \hline
\end{tabular}
\end{table}

For the performance evaluation, five APIs have been selected, encompassing read/write operations and public/private data, as detailed in Table~\ref{tab-apis}. The two data objects, i.e., \emph{policy} and \emph{ticket}, are stored as public and private data on the blockchain, respectively. The API \emph{Get a policy} retrieves a single policy, approximately a 1KB JSON object, based on the policy ID. \emph{Get all policies} retrieves all policies, totaling 91 policies and 56.97KB. \emph{Get all tickets} fetches all accessible tickets, which include eight tickets and amount to 5.67KB. It is noteworthy that read access control is enabled in the smart contract. Upon receiving the read request, the smart contract initially retrieves all valid \emph{readPolicy} according to the user and policy metadata, subsequently identifies accessible tickets, and returns them to \textsc{WebttCom}. \emph{Create a policy} generates a single \emph{writePolicy} of about 1KB, while \emph{Create a ticket} creates a single ticket, approximately 2KB, adhering to the write access control.

\begin{figure*}[!ht]
    \centering
    \subfigure[Web2 response time and Web3 processing time]{
    \begin{minipage}[t]{0.54\textwidth}
    \centering
    \includegraphics[width=\linewidth]{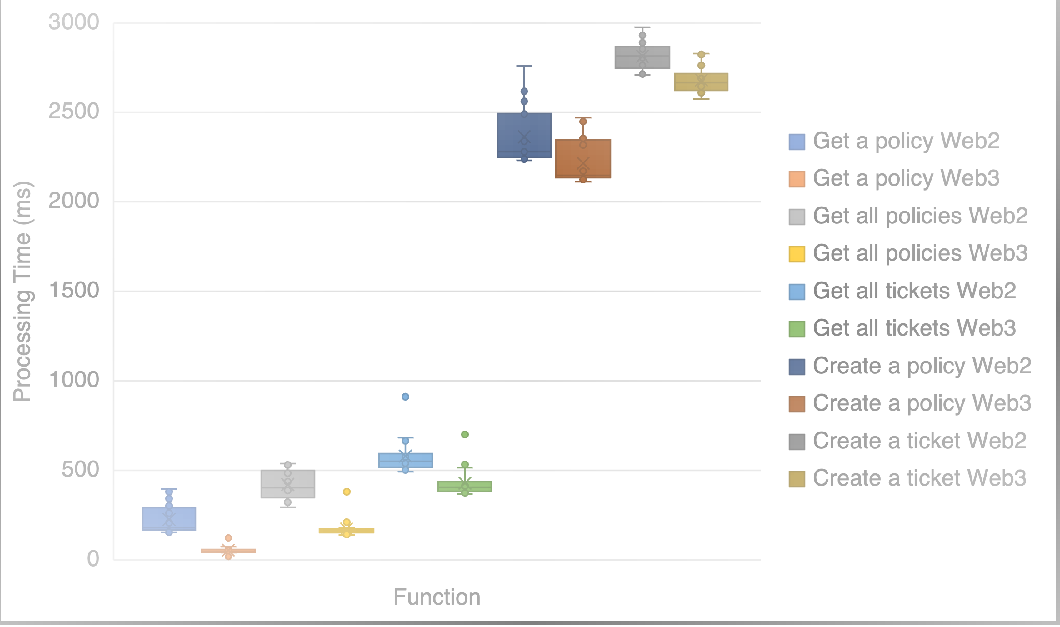}%2.4in
    \end{minipage}
    \label{fig:web23}
    }
    \subfigure[Processing time on \textsc{WebttCom}]{
    \begin{minipage}[t]{0.4\textwidth}
    \centering
    \includegraphics[width=\linewidth]{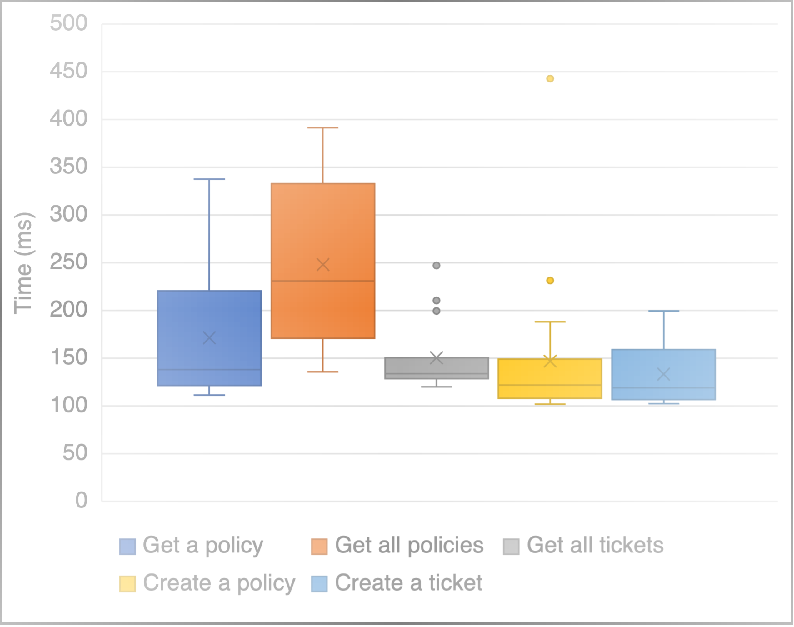}% 2.3in
    \end{minipage}
    \label{fig:overhead}
    }
    \vspace{-0.15in}
    \caption{Time performance of \textsc{WebttCom}}
    \label{fig:performance}
    \vspace{-0.15in}
\end{figure*}

Fig.~\ref{fig:performance} illustrates the overall response time and Web3 processing time of the APIs, labeled as Web2 and Web3, respectively. During the test, each API is executed independently 20 times in a sequence, with the overall response time derived from the HTTP response and the Web3 processing time extracted from the blockchain peer log. Therefore, the overall processing encompasses HTTP request handling in WebttCom, user authentication in \textsc{WebttCom}, transaction generation in \textsc{WebttCom}, communication between \textsc{WebttCom} and the blockchain, and Web3 processing time. As depicted in Fig.~\ref{fig:web23}, the three read APIs require less than 600 ms, and both the overall response and Web3 processing time increase with the complexity and data size of the applications. The two write APIs take approximately 2.5 seconds, attributable to the \emph{BatchTimeout} of 2 seconds, which could be expedited with a reduced \emph{BatchTimeout}. Operations on private data necessitate a longer processing time than those on public data. Fig.~\ref{fig:overhead} demonstrates the processing time on \textsc{WebttCom}, i.e., the discrepancy between the overall response time and Web3 processing time. \textsc{WebttCom} takes about 150 ms for the APIs, while \emph{Get all policies} takes roughly 200ms and exhibits the largest variance due to the substantial data transmission and resolution of a large number of policies. The evaluation results affirm that \textsc{WebttCom} can efficiently transition Web2 to Web3 applications.

\subsection{Survey Results}

We followed the same benchmarks in~\cite{9426788} by conducting a set of surveys to assess the \textit{Necessity}, \textit{Usability}, and \textit{Completeness} of the proposed framework \textsc{WebttCom} and the implementation. 

\subsubsection{Survey Questions}
\label{sec_questions}
The technical background and Web3 experience were asked, followed by the following questions from the domain experts to ensure the \textit{Necessity} of the new proposed framework \textsc{WebttCom}:
\begin{itemize}
\item What do you think of the pros and cons of Web2 and Web3 by now under your background?

\item What do you think of the necessity of a smooth transition between Web2 and Web3 (both Web2 to Web3 and Web3 to Web2) from your organizational and personal perspectives? 

\item What are the possible challenges during the transition that you think are required to be resolved immediately?
\end{itemize}

Further, we asked the following questions for feedback on \textsc{WebttCom} and overall \textit{Usability} and \textit{Completeness} of the implementation.

\begin{itemize}
\item To what extent does the Service Management System developed by UTS and BT, the implementation of the proposed framework \textsc{WebttCom}, match the principles of \textsc{WebttCom} and address the challenges above?

\item What are your suggestions to enhance the suitability of the new framework \textsc{WebttCom} and its implementation?
\end{itemize}

\subsubsection{Key Findings}
We summarized key findings from the surveys that can support the proposed framework \textsc{WebttCom} and implementation as a use case. 
Out of the 1,000 practitioners we invited, we received feedback from 58 valid respondents. Approximately 80\% of the responses were positive, and around 20\% were neutral; see Fig.~\ref{fig:survey}.
For clarity, we referenced five highly representative responses in Appendix-B during our discussion of the key findings, listed as follows:

\begin{figure*}[t]
    \centering
    \subfigure[Sentiment analysis regarding each category of responses]{
    \begin{minipage}[t]{0.48\textwidth}
    \centering
    \includegraphics[width=\linewidth]{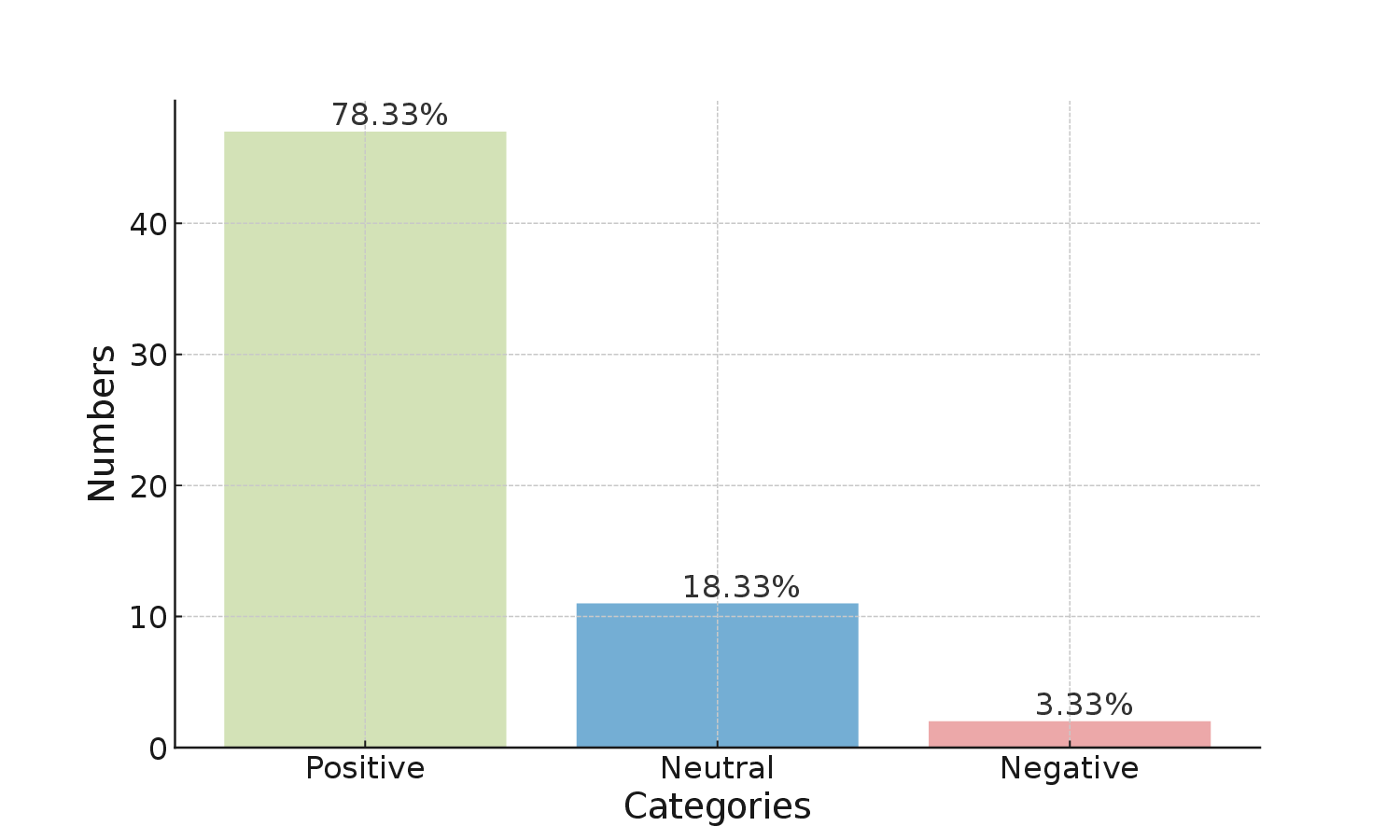}%2.4in
    \end{minipage}
    \label{fig:survey-bar}
    }
    \subfigure[Sentiment analysis regarding each response from different organizations]{
    \begin{minipage}[t]{0.48\textwidth}
    \centering
    \includegraphics[width=\linewidth]{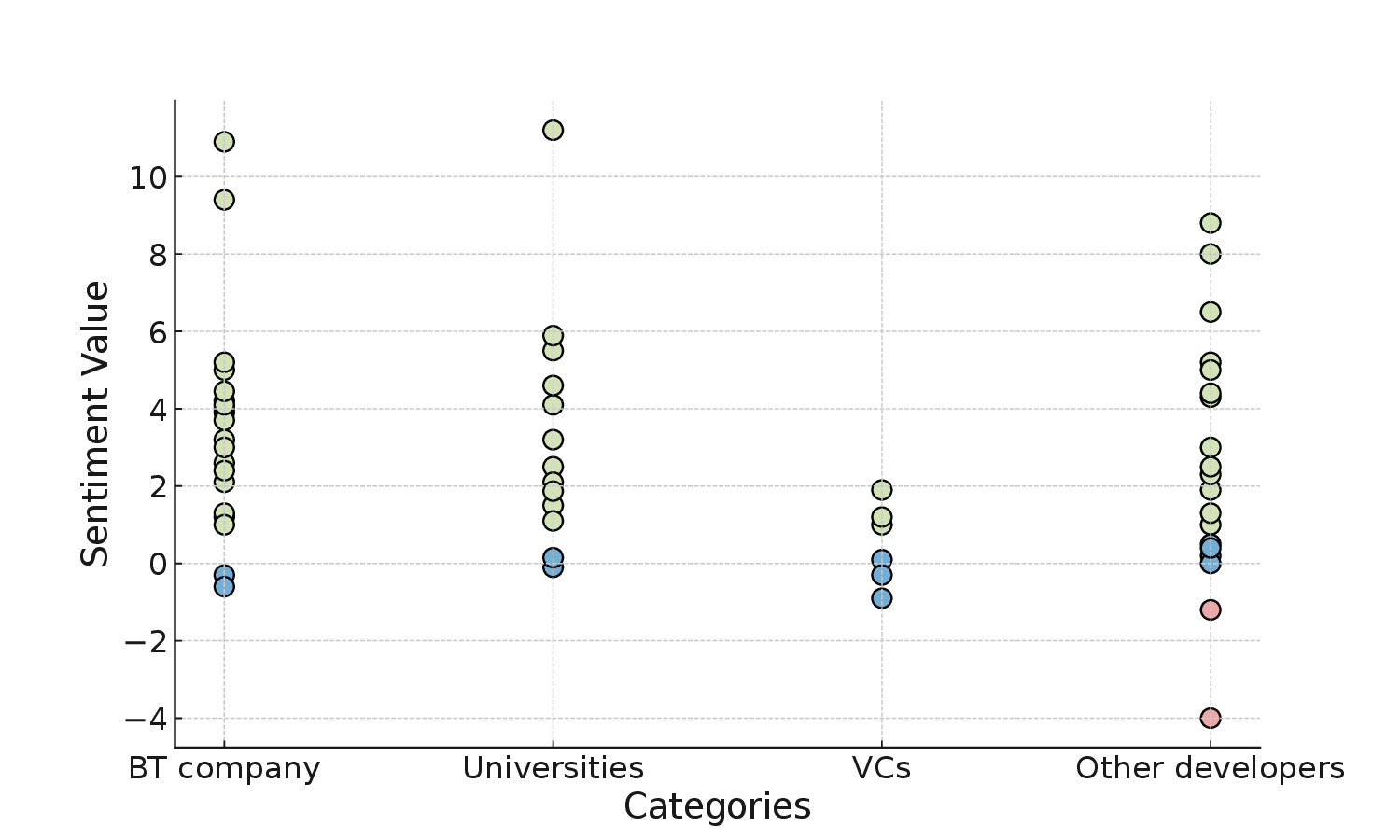}% 2.3in
    \end{minipage}
    \label{fig:survey-scatter}
    }
    \vspace{-0.15in}
    \caption{Sentiment analysis of the survey results}
    \label{fig:survey}
    \vspace{-0.15in}
\end{figure*}

\textit{1) \textbf{Necessity - enable guided and structured design: }}
Building Web3 technology into Web2 systems can solve critical security and trust issues in Web2 systems, especially for businesses across multi-organizations.
Web2 systems are centralized and can hardly achieve trust across organizations, while the inherent consensus mechanism of Web3 technology can provide verified single-ground truth and thus build trust across parties. 
The second surveyee stated, ``\textit{A transition from Web2 to Web3 will bring the data trustworthiness and cybersecurity guarantee from Web3 to Web2 systems. }''. 

It is important to have a smooth transition between Web2 and Web3. This is because Web2 is a mature technology, but developing Web3 applications could be challenging. As stated by the fourth surveyee with limited Web3 knowledge  ``\textit{The modification of the existing Web2 system should not be difficult. 
To avoid overloading the programmers with Web3 knowledge, they should not learn too much.}'', and the fifth surveyee stated, ``\textit{Web2 and Web3 should be able to coexist and interact smoothly.}'' Detailed challenges during the transition between Web2 and Web3 include 
\begin{itemize}
    \item The transition of two different technologies, as stated by the fourth surveyee ``\textit{Due to the differences in concept, technology, and tools between Web2 and Web3, it is difficult to integrate the Web2 system with the Web3 system.}''
    \item New access control schemes in Web3. The first surveyee stated ``\textit{Web3 is transparent, how to apply flexible and feasible access control becomes important for cases where data privacy is considered}'', and the third surveyee stated ``\textit{DLTs are a promising solution due to the ability of smart contracts to ensure that the required country-specific data management policies are agreed and enforced.}''
    \item Heterogeneous user management across Web2 and Web3. The first surveyee stated, ``\textit{Approach to apply the user management in a shared ledger may be a challenge as Web2 parties tend to have separate user management systems locally.}''
    \item High development and transition cost. The first and the fifth surveyees stated ``\textit{Easing the transition by using automated tools is normal in Web2 applications and is also essential during the transition between Web2 and Web3.}'' and ``\textit{Challenges include lack of experienced developers, lack of available development resources, tools.}''
\end{itemize}

\textit{2) \textbf{Usability - flexible access control: }}
The developed flexible access control can implement access control policies as designed.
The service provider can define access control policies, including data storage and write and read policies. The smart contract on the Hyperledger Fabric enforces all the policies and controls access to on-chain data.
All surveyees have confirmed the developed access control mechanism.
The second surveyee stated, ``\textit{The system can enforce all the expected data governance and access control policies.}'', and the third surveyee stated ``\textit{In evaluating the solution we have demonstrated that a DLT Hyperledger layer can meet the required success criteria related to cross-country access control and user-management; as well as connecting to a traditional SaaS workflow management layer.}''

\textit{3) \textbf{Usability - compatible user management: }}
The developed user management compatibly manages Web2 and Web3 users. When a user submits a Web2 registration request, Web2 and Web3 accounts are simultaneously created and managed by the developed framework. When users log in and submit requests, the framework can perform authentication with the users' Web2 credentials and process Web3 requests and responses in the representation of the users.
All chosen surveyees are satisfied with the developed user management. The second and fourth surveyees stated, ``\textit{A user can use one identity to access Web2 and Web3 services.}'' and ``\textit{Web2 programmers are not required to touch too much.}''

\textit{4) \textbf{Usability - easy integration with existing platforms and services: }}
The developed system seamlessly integrates with existing service management on ServiceNow. All operations are conducted in ServiceNow, including login, access control policy management, and service ticket management. The workflow remains the same with the Web2 version. Nevertheless, all data are securely saved on the Hyperledger rather than disconnected databases and can be verified with the Web3-certified data hash.
As the second surveyee stated, ``\textit{Users can access Web3-certified data services from the SaaS. The complicated Web3 details are transparent to users.}''

\textit{5) \textbf{ Usability - highly automated development tools: }}
Highly automated development tools are implemented to generate and document OpenAPI-compliant REST APIs of both Web2 and Web3 registries. This significantly improves programming efficiency by sharing the same development workflows and schema formats between Web2 and Web3. As the 1st, 2nd, and 4th surveyees gave positive feedback by stating ``\textit{Automated development tools significantly reduce the programmers' workload and has almost become a must-use tool during the current workflow.}''.

\textit{6) \textbf{Completeness - sufficient decentralization with robust data privacy and governance: }}
The surveyees satisfy the framework \textsc{WebttCom} and the use case regarding achieving sufficient decentralization level while ensuring strong data privacy and governance; as the first surveyee stated ``\textit{The framework appears to be covering most perspectives including the access control, data privacy, data governance, user management, connections to existing SaaS, and development productivity.}''. This is done by successfully applying the flexible access control in a decentralized manner, i.e., smart contracts on the Hyperledger Fabric. Robust data privacy and governance can thus be offered to ensure secure and transparent data management while fostering trust among participants in a distributed and tamper-resistant manner.

\textit{7) \textbf{Completeness - limited cases: }}
The fourth and fifth surveyees stated ``\textit{More commercial Web2 systems should be integrated with Web3.}'' and ``\textit{Need to find suitable business use cases to demonstrate the benefit of Web3}''.
Three surveyees noted that the proposed framework \textsc{WebttCom} requires more business use cases to support by potentially integrating the existing commercial Web2 projects. While we admit this limitation at the time of writing, this work originates from solving the transition issue from Web2 to Web3, which came across by our business partner proceeding with their latest strategy. This work is new and lacks sufficient existing use cases. Therefore, this work focuses on demonstrating how \textsc{WebttCom} solves the research question, and we postpone extending use cases to improvements and future works.

\textit{8) \textbf{Potential Improvements: }}
Among the chosen representative respondents, four indicated that the framework and the provided use case seem limited to permissioned blockchain platforms such as HyperLedger Fabric. They suggested implementing this framework on other prominent platforms, such as Ethereum, and with more advanced scalability solutions such as blockchain sharding~\cite{8954616,Yu2020,10.1007/978-3-031-23495-8_6} and cross-chain designs~\cite{yu2022crosschain}, necessitates additional testing. Another highlighted respondent noted that while the integration with ServiceNow is commendable, there's a need to further test its compatibility with services and functionalities offered by Platform as a Service (PaaS) or Infrastructure as a Service (IaaS) platforms, such as AWS and Azure. Notably, these representative insights align with the general sentiment expressed in the 58 valid responses we received.

%% file: sections/Discussion.tex
\section{Limitations and Validity}\label{sec:discussion}

We apply the guidelines \cite{wohlin2012experimentation} to discuss key threats to the validity (\textit{construct}, \textit{internal}, \textit{external}) of this work.

\textit{\textbf{Construct validity}} reflects what extent the research questions and methodology are appropriately used in a study. A threat in Stage 3 (i.e., the surveys) is whether or not our surveyees are representative. To reduce this threat, we invited 1,000 practitioners who come from the company, universities, and VC members
who are applying for our Web3 implementation. In addition, the areas that the participants have worked on cover a wide range of domains (e.g., blockchain and AI applications). However, they may not be representative of all practitioners, and not all the intended surveyees are expected to respond in time. 
To mitigate this potential bias, we have carefully chosen questions and topics. Our survey respondents completed the surveys based on their opinions and perceptions. It is possible that they conflate the skills that are very important and the skills that are very relevant to their projects or industrial contexts.

\textit{\textbf{Internal validity}} focuses on factors that may influence the validity of the results. The main
threat in our study is whether the study process we designed and the Web3 framework and application that we proposed can answer our research questions.  It is also possible that we draw the wrong conclusions about respondents’ perceptions from their
comments. To minimize this threat, we read transcripts many times and checked the survey results
and the corresponding comments several times.

The selection of statements produced at the end of the surveys may not be comprehensive and may be biased to the background of experts—who may not be able to articulate their own opinions. To mitigate this bias, we have taken the following steps:
\begin{itemize}
\item Aside from asking direct questions about their opinions about Web3 and applications that we designed and implemented, we also asked them to discuss general topics that they had not explicitly mentioned. The topics were selected from Web3 textbooks and online resources; they include concepts, comprehension, programming language, requirements, design implementation, testing, and tool usage.
\item Three authors have performed data analysis to cross-check their answers by using card sorting \cite{fincher2005making}, and we carefully examined and only included relevant statements. 
\end{itemize}

\textit{\textbf{External validity}} concerns the generality of our study results to other settings. Our results and summaries are based on the Web3 framework and application and survey participants’ opinions instead of a rigorous analysis of participants' claims. It is possible that opinions regarding the Web3 framework and application differ from one participant to another. To improve the generalizability of our results, we invited 1,000 practitioners and obtained 58 respondents. Still, our findings may not generalize or represent the perception of all software engineers. For example, the respondents are from one company and VCs that are interested in Web3 technologies and closed universities. 
It can be noticed in Fig.~\ref{fig:survey-scatter} that a sentiment bias in certain categories. The framework was initially designed for BT Company, leading to skewed positive feedback. A similar bias exists in the Universities category, potentially due to some participants being affiliated with the authors' institution. In contrast, feedback from VCs and other developers, sourced from GitHub and mailing lists, appears more balanced and free from conflicts of interest.
It would be interesting to perform another study to investigate more software engineers to perceive the future benefits and limitations of the Web3 framework and application.

\textit{\textbf{Platform validity}} refers to the extent to which the proposed framework can be universally applied across multiple blockchain platforms. While our design principle for the blockchain layer is inherently blockchain-agnostic, our case study predominantly used Hyperledger for illustration. This choice was primarily to demonstrate the capabilities of our framework in a clear and specific manner. However, it is essential to stress that the architecture and design of our framework, \textsc{WebttCom}, do not restrict its applicability. The core principles and mechanisms of \textsc{WebttCom} can be effectively translated to other blockchain platforms with the necessary configuration adjustments. The foundation remains consistent, versatile, and designed for broad adaptability.

To elucidate further on the adaptability of \textsc{WebttCom} with other platforms such as Ethereum: The crux of our framework lies in its modular design. The blockchain layer, which is designed to be blockchain-agnostic, can interface with Ethereum's smart contract system just as it does with Hyperledger. Ethereum, with its robust smart contract capabilities, can effectively replace the smart contract for data provisioning and the modules for write, update, and read controls present in our framework. Furthermore, Ethereum's dynamic storage system can be leveraged to handle service data storage in our blockchain layer, substituting for private and public data repositories. The backend interpreter, which serves as the bridge between Web2 and Web3, remains consistent. Its Web2-Web3 transition module can be configured to forward requests to Ethereum's smart contracts, ensuring seamless communication. Similarly, user management for Web2 and Web3 users in the backend would work harmoniously with Ethereum, ensuring an integrated user experience.

%% file: sections/RelatedWork.tex
\section{Related Work}\label{sec:related}

%{\color{blue} (The related work should include 1) In software engineering, any existing papers studying the Web2/web3 connection/transition pattern); 2) Existing Web2/web3 connection/transition solutions that comply or not comply with our proposed pattern. If complied, pointing out the lack of pattern, otherwise pointing our their weaknesses; 3)...)}
%access

In this section, we give a quick overview of the notion of Web3 and then introduce ways of building DApps on top of blockchains.

\smallskip
\noindent\textbf{Notion of Web3.}
The concept of Web3 was first proposed by Wood~\cite{woodweb3} with an initial discussion focusing on blockchain-based digital infrastructure. Later, Weyl et al. \cite{weyl2022decentralized} illustrate the ways of building a decentralized society by exploring Web3-related applications, requirements, opportunities, and challenges.  Wang et al.~\cite{wang2022exploring} provide the discussion between Web3 and blockchain from the perspective of architecture design. The work has identified a total of 12 types of designs according to the data workflow of access, computation, and storage.  A series of reports from Consensys \cite{repoconsys} investigates the economic performances created by decentralized networks. Web3 economy covers many on-chain DeFi protocols \cite{jiang2023decentralized} such as stablecoins, borrowing, lending, and leverage, distributed governance protocols (e.g., DAO)~\cite{yu2023leveraging}\cite{wang2022empirical} and many innovations combined with other technologies~\cite{yu2023predicting,wang2023referable}. These protocols are built on top of smart contracts.  Liu et al. \cite{liu2021make} explore three types of infrastructural enablers, including the single smart-contract powered chain, federated contracts, and interoperable blockchain platforms. However,  Web3 sofar still confronts high-level controversies in terms of its definition and application. We, in this paper, extend our exploration by complying with its core \textit{decentralization} nature guaranteed by underlying blockchain services.
 
\smallskip
\noindent\textbf{Constructing Web3 Applications.} Traditional Web1 and Web2 applications rely on centralized backend servers for computations and storage. Web3 applications \cite{wang2022exploring} replace these servers with decentralized blockchain platforms \cite{wood2014ethereum}\cite{bsc}. Building a Web3 application requires three phases: (i) \textit{embedding the wallet for access}, (ii) \textit{connecting frontend with blockchain platforms}, and (iii) \textit{operating executions on-chain}. A wallet~\cite{chatzigiannis2021sok}\cite{karantias2020sok} is backed by locally running lightweight nodes and helps Web3 users to create on-chain accounts (in the form of \textit{address} \cite{bonneau2015sok}) as their identities. A user can initiate a request by sending a transaction from wallets to the transaction pool. Smoothly processing the received requests from the front end requires a suite of standard protocols, including token standards (e.g., EIP \cite{eip}, BEP \cite{bep} or even BRC~\cite{wang2023understanding}) and unified APIs. All the compiled bytecodes will be executed on-chain by iterative state transitions. The consensus mechanism is critical in maintaining state consistency and chain stability~\cite{garay2015bitcoin}. In some cases, external techniques are needed for supportive functionalities such as distributed storage~\cite{benet2014ipfs}, layer-two computations~\cite{gudgeon2020sok}, cross-chain bridges~\cite{zamyatin2021sok}\cite{wang2023exploring} or on-chain oracles~\cite{breidenbach2021chainlink}. Our work achieves more than building a simple DApp that is independent of operating blockchains. We, instead, develop a general interpreter to connect Web2 applications to the current leading Web3 platforms from the access layer to the chain layer. This gives an educational study for the community.

%% file: sections/Conclusion.tex
\section{Conclusion}\label{sec:conclusion}

In this paper, we gave a research question by exploring the connotation of Web3 and the key differences between Web2 and Web3 applications. We proposed a new framework, named \textsc{WebttCom}, to enable a seamless transition from Web2 to Web3. In particular, \textsc{WebttCom} can smoothly connect traditional Web2 applications to mainstream Web3 blockchain platforms while additionally ensuring high data privacy and governance, and improving development productivity. Our design innovative introduces an interpreter mechanism that can aggregate and deal with requests between Web2 and Web3 spaces.  Accordingly, we implemented a full-stack system consisting of over 11,351 lines of code, conducted a performance evaluation, and launched a survey to assess the effectiveness of our framework. Corresponding surveys with experienced participants confirmed \textsc{WebttCom} satisfies our research question with a few possible limitations of the framework and its related business cases. We further provide our recommended improvements, including the extension of the framework to involve more blockchain platforms and search for more business cases.

% such as the connection with AWS and Azure.

%% file: sections/Appendix-1.tex
\subsection*{APPENDIX-A}
The entrance of the implemented service management contract is shown in Listing~\ref{code_contractentry}. The service management contract extends the Hyperledger Fabric \textit{Contract} class. Users can invoke the functions by sending transactions containing the function names and parameters. Users can \textit{create}, \textit{update}, and  \textit{readWithFilter} objects saved on the Hyperledger Fabric. Functions are implemented in controller files. 
\begin{lstlisting}[style=JavaScript, caption={Entry Point of the Ticket Management Contract.},label={code_contractentry},basicstyle=\ttfamily\scriptsize]
import { Context, Contract, Info, Returns, Transaction } from 'fabric-contract-api';
import { Controller } from 'web3-controller';

export class ServiceContract extends Contract {
    @Transaction(true)
    @Returns('string')
    public async create(ctx: Context, attributes: string, publicPayload:string): Promise<string> {
        return await new Controller(<'policy'|'object'|'user'>).createRegistry(ctx, ctx.stub.getTxID(), attributes, publicPayload);
    }    
    @Transaction(true)
    @Returns('string')
    public async update(ctx: Context, dataId: string, attributes: string, publicPayload:string): Promise<string> {
        return await new Controller(<'policy'|'object'|'user'>).updateRegistry(ctx, dataId, attributes, publicPayload);
    }
    @Transaction(false)
    @Returns('string')
    public async readWithFilter(ctx: Context, publicPayload:string): Promise<string> {
        return await new Controller(<'policy'|'object'|'user'>).readRegistryWithFilter(ctx, publicPayload);
    }
}
\end{lstlisting}

The read and update policy checking is shown in Listing~\ref{code_readcheck}. The function checks all policies linked to the corresponding objects, which are given by the parameter \textit{policyIDs}. The function fetches the policies with the \textit{policyIDs}, checks \textit{user}, \textit{location}, \textit{validTime}, \textit{updateTime}, and \textit{priority}, and returns the effective policy.

\begin{lstlisting}[style=JavaScript, caption={Read and Update Policy Checking Function.},label={code_readcheck},basicstyle=\ttfamily\scriptsize]
    public async getActiveReadAndUpdatePolicy(ctx: Context, location: string, policyIDs: string[]): Promise<any> {
        let effectivePolicy: any = {};
        let user = lla.getUserInfo(ctx);
        let ts = ctx.stub.getTxTimestamp().getSeconds() * 1000;
        let latestTimestamp = 0;
        let highestPriority = 0;
        for (let policyID of policyIDs){
            let policyObj: any = await new pc.policy().loadPolicyFromID(ctx, policyID);
            let updateTime = await lla.getLastUpdateTime(ctx,policyID);
            let permittedUser = new userInfo(policyObj.privilege.rowPolicy.userAffiliation)
            if (user.match(permittedUser)) {
                if (location == policyObj.privilege.rowPolicy.location.countryName) {
                    let startStamp = new Date(policyObj.validFor.startDateTime).valueOf();
                    let endStamp = new Date(policyObj.validFor.endDateTime).valueOf();
                    if (ts >= startStamp && ts <= endStamp) {
                        if (policyObj.priority > highestPriority) {
                                effectivePolicy = policyObj;
                                latestTimestamp=updateTime;
                                highestPriority=policyObj.priority;
                        }
                        if (policyObj.priority == highestPriority) {
                            if (updateTime > latestTimestamp) {
                                effectivePolicy = policyObj;
                                latestTimestamp=updateTime;
         }}}}}}
        return effectivePolicy;
    }
\end{lstlisting}

The Web2-Web3 interpreter is shown in Listing~\ref{code_interpreter}.
When receiving a Web2 request, the interpreter creates a Hyperledger Fabric connection, including the connection profile, Fabric user, channel name, and contract name, according to the payload of the Web2 request. The interpreter then forwards the Web2 request and parameters to the corresponding functions in the Web3 contract. When receiving Web3 results, the interpreter returns them to the Web2 request.

\begin{lstlisting}[style=JavaScript, caption={Web2-Web3 Interpreter.},label={code_interpreter},basicstyle=\ttfamily\scriptsize]
    export async function Trans(
      _fabric: ITransient, context?: Required<Pick<IUserManagement, "userID" | "userPWD">>, fnc: string, ...args: string[]
    ): Promise<string> {
      let promise: Promise<string> = new Promise(async (resolve, reject) => {
        try {
          const fabricConCof: FabricConfiguration = new FabricConfiguration(
            _fabric.connection,     /* The Fabric connection profile */
            _fabric.walletPath,     /* The path to Web3 credentials */
            _fabric.walletID,       /* The Web3 user */
            _fabric.channelName,    /* The Fabric channel name  */
            _fabric.scName,         /* The smart contract name */
            _fabric.listen          /* Fabric event listener*/
          );
          const [gateway, connectionProfile, connectionOptions] = await getFabricConfig(context, _fabric);
          await gateway.connect(connectionProfile, connectionOptions);
          let network = gateway.getNetwork(fabricConCof.channelName);
          let contract = (await network).getContract(fabricConCof.scName);
          const channel = (await network).getChannel();
          let result: Buffer | any;
          result = await contract.submitTransaction(fnc, ...args);
          await gateway.disconnect();
          if (result) { resolve(result.toString()); }
          else { resolve("Invoking ${fnc} succeeds."); }
        } catch (error) { reject(error); }
      });
      return promise;
    }
\end{lstlisting}

%% file: sections/Appendix-2.tex
\subsection*{APPENDIX-B}\label{subsec:appendix-b}
In the pursuit of gathering comprehensive insights on the topic, we reached out to a diverse group encompassing 1,000 individuals and institutions through various channels including GitHub, mailing lists, corporations, and academic universities. Despite this extensive outreach, we received a total of 58 valid responses by the stipulated deadline. While the response rate might appear limited given the vast initial pool, it's imperative to underscore the rigor and validity of the data collected. Every response underwent meticulous validation to ensure its relevance and accuracy. Out of these valid contributions, five responses were handpicked to be showcased in this appendix due to their exemplary representativeness of the collective sentiments and insights. These five chosen responses stand as testament to the varied perspectives and depth of knowledge within our respondent pool, further bolstering the credibility of our methodologies and the consequent findings.

\begin{tcolorbox}[%
title = {First response},
    enhanced, 
    breakable]
\textbf{1.	What is your job in the organization? Technical or non-technical?}

I am a software developer in my organization.
\\ \hspace*{\fill} \\

\textbf{2.	What is your Web3 background?}

I have 4-year developing experience on Web3 (expert)
\\ \hspace*{\fill} \\

\textbf{3.	What do you think of the pros and cons of Web2 and Web3 at present? }

Web2:

Pros: mature, design patterns, tools, and workflows have been thoroughly explored and developed, and have been proven applicable for many huge and complicated tasks.

Cons: heavily relies on trustworthy centralized entities. Building trust among parties heavily relies on social engineering, which often causes interests dispute.

Web3:

Pros: achieve trustworthy relationships among parties without centralized entities in a decentralized manner.

Cons: immature, lacks design patterns, tools, and workflows. Popularity is not as good as Web2. 
\\ \hspace*{\fill} \\

\textbf{4.	What do you think of the necessity of a smooth transition between Web2 and Web3 from your organization and your personal perspectives? Web2 to Web3 and Web3 to Web2?}

It is of importance to have a smooth transition between Web2 and Web3. 

Web2 project sometimes not only needs to focus on the trust issue but also needs to have an efficient incentive mechanism to encourage the public to use the services. And Web3 fits the requirement by applying decentralized ledger technology.

Many Web3 applications have proven their potential in many areas including finance, asset trading, gaming, law… However, Web3 is still new to the market and needs to increase its popularity by collaborating with those giant companies and their mature products.
\\ \hspace*{\fill} \\

\textbf{5.	What possible challenges during the transition do you think are required to be resolved?}

\begin{enumerate}
\item Web3 is transparent, and how to apply flexible and feasible access control becomes important for cases where data privacy is considered.
\item Approach to applying user management in a shared ledger may be a challenge because in Web2 parties tend to have separate user management systems local.
\item How to connect with existing mature products, such as SaaS, may be a challenge.
\item Easing the transition by using automated tools is normal in Web2 applications and is also essential during the transition between Web2 and Web3. Having such tools that can also be seamlessly integrated into existing Web2 and Web3 frameworks may be a challenge.
\\ \hspace*{\fill} \\
\end{enumerate}

\textbf{6.	To what extent the Service Management System developed by UTS and BT UK matches the framework and solve the questions?}

a.	Access control 

Satisfied, the data-driven policy is smart. 

b.	User management 

Satisfied

c.	Connection with SaaS 

Satisfied with the combination with ServiceNow which has been widely used in many existing products.

d.	Automated development tools

TSOA is an interesting tool as Typescript is very popular and the OpenAPI standard has almost been a must-use tool during the current workflow in Web2 development.
\\ \hspace*{\fill} \\

\textbf{7. What are your suggestions to enhance the suitability of the new framework \textsc{WebttCom} and its implementation?}

The framework appears to be covering most perspectives including access control, data privacy, data provenance, user management, connections to existing SaaS, and development productivity. And the implementation also shows a good match to the framework. One point that could be improved is the extension of Web3 platforms where other platforms can be involved other than HyperLedger. Then any possible changes to the implementation to match the framework should be considered.
\\ \hspace*{\fill} \\
\end{tcolorbox}

\begin{tcolorbox}[%
title = {Second response},
    enhanced, 
    breakable]
\textbf{1.	What is your job in the organization? Technical or non-technical?}

I am a researcher and software developer in my organization.
\\ \hspace*{\fill} \\

\textbf{2.	What is your Web3 background?}

I have 4-year research and development experience on Web3.
\\ \hspace*{\fill} \\

\textbf{3.	What do you think of the pros and cons of Web2 and Web3 at present? }

Web2:

Pros: Web2 technology has been developed for decades. There are many mature solutions for various requirements and development tools for rapid development and simple management. Consumers have been well educated on the Web2 service pattern. 

Cons: The biggest challenge of Web2 systems is the trust issue. For a single Web2 service, consumers have to trust Web2 service providers, and the service providers need to trust developers, infrastructure providers, etc. When multiple Web2 systems are connecting to each other, they need to trust the services and data from others. In the case of untrusted relations or inconsistent data, it is challenging to solve disputes and provide services as a whole.

Web3:

Pros: Web3 can provide single ground truth to all participants in a trustless way. Different parties do not have to build trust with each other.

Cons: Web3 is a new technology. The architecture and process of Web3 services are very different from those of Web2. Deploying and maintaining Web3 infrastructure and applications could be very time-consuming and risky. Consumers need to be educated about changes from Web3.
\\ \hspace*{\fill} \\

\textbf{4.	What do you think of the necessity of a smooth transition between Web2 and Web3 from your organization and your personal perspectives? Web2 to Web3 and Web3 to Web2?}

We are aware of the benefit of Web3 and have plan to employ Web3 technology in our business to provide trusted services and reduce business loss. 

A transition from Web2 to Web3 will bring the data trustworthiness and cybersecurity guarantee from Web3 to Web2 systems. The Web3 token mechanism can support incentive schemes in Web2 applications.

The smooth transition from Web3 to Web2 applications will promote Web3 technology and applications. There have been many Web3 applications, like tokens and contracts. However, Web3 applications are only popular among the Web3 community because the requirements and access to Web3 applications are very different from widely-used Web2 applications and can be hard for general users. A Web3 to Web2 transition will simplify access and encourage more users to try Web3 technology and applications.
\\ \hspace*{\fill} \\

\textbf{5.	What possible challenges during the transition do you think are required to be resolved?}

1)	How to seamlessly integrate Web3 services into existing Web2 systems

2)	How to implement data governance and privacy policies in Web3

3)	How to manage Web2 and Web3 user identities simultaneously

4)	How to reduce integration and development cost

5)	How to efficiently manage Web3 infrastructure

6)	Can Web3 technology support large-scale applications
\\ \hspace*{\fill} \\

\textbf{6.	To what extent the Service Management System developed by UTS and BT UK matches the framework and solve the questions?}

a.	Access control 

Satisfied. The system can enforce all the expected data governance and access control policies. 

b.	User management 

Satisfied. A user can use one identity to access Web2 and Web3 services. 

c.	Connection with SaaS 

Satisfied. Users can access Web3-certified data services from the SaaS. The complicated Web3 details are transparent to users.

d.	Automated development tools

Satisfied. The automatic document generation technology can save 20% on development. The generated API documents are accurate and give execution samples.
\\ \hspace*{\fill} \\

\textbf{7. What are your suggestions to enhance the suitability of the new framework \textsc{WebttCom} and its implementation?}

The developed WebttCom is a good start for transiting Web2 to Web3. I suggest the team develop more service functions to the framework, such as managing ticket files across multiple departments and organizations, such that the system usability could be improved. The framework can also be integrated with other Web2 businesses, such as workflow management. Besides new functions, comprehensive trials need to be conducted to verify the stability and capacity.
\end{tcolorbox}

\begin{tcolorbox}[%
title = {Third response},
    enhanced, 
    breakable]
Hence, there are  two options for the paper: 1) The survey section could provide an aggregated view from the project team’s view on how well the prototype addresses the needs of Web3, without attributing views to individual organizations. Or 2) On the BT side we could write a section for the evaluation part of the paper.
 
To give you a flavor of what a BT-authored section would look like: It would explain that Web3 is not well defined at the moment and is currently often linked to ideological aims around openness and decentralization, compared to the current internet which is seen, by the Web3 community,  as dominated by “big tech”. The Web3 concept is a very nascent area at the moment where it will be interesting to see how it evolves and fits with the current Web2 world in which we live.
 
Web3 needs to answer a number of core questions related to complexity, scalability cost; and, critically, also cost-effectiveness and accountability, which are often more important to commercial organizations than adherence to decentralization principles. However, the lack of well-formed answers to these high-level generic considerations should not prevent us from identifying use cases and prototyping solutions to further our knowledge of the opportunities.
 
Looking at the Web3 technology portfolio it already includes some useful enabling technologies, including distributed ledger technologies (DLTs) which, with their inherently decentralized architecture, can be useful for specific applications. On their own they are unlikely to form the universal technology base for the next generation of the Internet; however, there are real use cases that exist today where DLTs may have an important role to play.  The work with UTS is an example of this and examines a real-world use case related to IT service management across different country boundaries where different data privacy regulations apply.
 
In this context, DLTs are a promising solution due to the ability of smart contracts to ensure that the required country-specific data management policies are agreed upon and enforced. For the current prototyping project, it was assumed that the transition to a DLT-based solution will require interworking with the more traditionally architected solutions which are already in common usage in commercial ecosystems. In evaluating the solution we have demonstrated that a DLT Hyperledger layer can meet the required success criteria related to cross-country access control and user management; as well as, connecting to a traditional SaaS workflow management layer.
 
In meeting these criteria the prototype has demonstrated sufficient promise that we are examining the next stage in research and development, which is to run test scenarios, utilizing realistic commercial data flows, through the prototype. This measured exploration and transition to DLT technologies show the path which we believe many established commercial organizations will follow as they gain confidence that the decentralized approach can meet the required business requirements. The approach of monitoring the evolving Web3 concepts, and engaging with the core technology developments, to experiment and build early insight on use cases for these emerging capabilities (e.g. DLT, smart contracts) is a pragmatic approach for ensuring that value can be gained.
\end{tcolorbox}

\begin{tcolorbox}[%
title = {Fourth response},
    enhanced, 
    breakable]
\textbf{1.	What is your job in the organization? Technical or non-technical?}

I am a software developer in my organization. Technical.
\\ \hspace*{\fill} \\

\textbf{2.	What is your Web3 background?}

I have a half year of Web3 background.  My Web3 knowledge is limited.
\\ \hspace*{\fill} \\

\textbf{3.	What do you think of the pros and cons of Web2 and Web3 at present? }

Web2:

Pros: Web2 has a large development community including programmers, tools, and solutions. Web2 has been widely applied to many organizations and it has been verified as a stable technology.

Cons: In the case of multiple organizations, it is not used for cross-validation and trust in the scenario.
\\ \hspace*{\fill} \\

\textbf{4.	What do you think of the necessity of a smooth transition between Web2 and Web3 from your organization and your personal perspectives? Web2 to Web3 and Web3 to Web2?}

Since most systems are based on Web2, we must focus on the transition from Web2 to Web3.

The modification of the existing Web2 system should not be difficult.

In order to avoid overloading the programmers with Web3 knowledge, they should not learn too much.

It is also possible to apply the existing Web2 technology to Web3-based systems.

It is not necessary for normal users to recognize the transition.
\\ \hspace*{\fill} \\

\textbf{5.	What possible challenges during the transition do you think are required to be resolved?}

1)	Because of their limited understanding of Web3, programmers are reluctant to transfer their existing Web2 systems to Web3.

2)	Due to the differences in concept, technology, and tools between Web2 and Web3, it is difficult to integrate the Web2 system with the Web3 system.
\\ \hspace*{\fill} \\

\textbf{6.	To what extent the Service Management System developed by UTS and BT UK matches the framework and solve the questions?}

a.	Access control 

Solved. Web3 implements access control, which means that Web2 programmers are not required to touch too much.

b.	User management 

Solved. It’s similar to the above. Web3 implements access control, which means that Web2 programmers are not required to touch too much.

c.	Connection with SaaS 

Solved, the modification of the existing Web2 system is not too extensive. It is also possible to apply the existing Web2 development method and architecture to the transition.

d.	Automated development tools

Solved, automated development tools significantly reduce the programmers' workload.
\\ \hspace*{\fill} \\

\textbf{7. What are your suggestions to enhance the suitability of the new framework \textsc{WebttCom} and its implementation?}

Attempts to integrate the existing commercial Web2 system's functionality with Web3.

More commercial Web2 systems should be integrated with Web3.

Provides more Web3 services.
\end{tcolorbox}

\begin{tcolorbox}[%
title = {Fifth response},
    enhanced, 
    breakable]
\textbf{1.	What is your job in the organization? Technical or non-technical?}

I am a developer, technical
\\ \hspace*{\fill} \\

\textbf{2.	What is your Web3 background?}

3 years of exposure to Web3 development
\\ \hspace*{\fill} \\

\textbf{3.	What do you think of the pros and cons of Web2 and Web3 at present? }

Web2: pros – mature technology, plenty of resources available; cons – being a centralized system, trust can be an issue
Web3: pros – has built-in trust mechanisms; cons – new tech, not many resources for developers
\\ \hspace*{\fill} \\

\textbf{4.	What do you think of the necessity of a smooth transition between Web2 and Web3 from your organization and your personal perspectives? Web2 to Web3 and Web3 to Web2?}

Certain businesses require the trusted services provided by web3, but not all. Many existing services are fine with Web2. Only services that deal with multi-organizations require web3.
Web2 and web3 should be able to coexist and interact smoothly.
\\ \hspace*{\fill} \\

\textbf{5.	What possible challenges during the transition do you think are required to be resolved?}

Challenges include a lack of experienced developers, a lack of available development resources, and tools.
\\ \hspace*{\fill} \\

\textbf{6.	To what extent the Service Management System developed by UTS and BT UK matches the framework and solve the questions?}

a.	Access control 

Completed.

b.	User management 

Completed.

c.	Connection with SaaS 

A good start in connection with ServiceNow

d.	Automated development tools

Some tools were developed.
\\ \hspace*{\fill} \\

\textbf{7. What are your suggestions to enhance the suitability of the new framework \textsc{WebttCom} and its implementation?}

Need to find suitable business use cases to demonstrate the benefit of web3.
\end{tcolorbox}